\documentclass[ALICE,manyauthors]{cernphprep}

\usepackage{amsmath}
\usepackage{bm}

\usepackage[comma,square,numbers,sort&compress]{natbib}
\usepackage{color}
\usepackage{hyperref}

\usepackage{lineno}
\usepackage{multirow} 
\usepackage{ltxcmds}

\usepackage{placeins}
\usepackage{subfigure}
\usepackage{calrsfs}

\usepackage[T1]{fontenc}
\usepackage{orcidlink}

\begin{document}%

\begin{titlepage}
\PHyear{2025}       
\PHnumber{237}      
\PHdate{13 October}  

\title{Space-time evolution of particle emission in p--Pb collisions at {\ensuremath{\pmb{\sqrt{s_{\rm NN}}}}}{\bf ~=~5.02}~TeV with 3D kaon femtoscopy}
\ShortTitle{Space-time evolution of particle emission in p--Pb collisions}

\Collaboration{ALICE Collaboration\thanks{See Appendix~\ref{app:collab} for the list of collaboration members}}
\ShortAuthor{ALICE Collaboration} 

\begin{abstract}
The measurement of three-dimensional femtoscopic correlations between identical charged kaons (K$^\pm$K$^\pm$) produced in p--Pb collisions at center-of-mass energy per nucleon pair $\sqrt{s{_{\rm NN}}} = 5.02$~TeV with ALICE at the LHC is presented for the first time.
This measurement, supplementary to those in pp and Pb--Pb collisions, allows understanding the particle-production mechanisms at different charged-particle multiplicities and provides information on the dynamics of the source of particles created in p--Pb collisions, for which a general consensus does not yet exist. It is shown that the measured source sizes increase with charged-particle multiplicity and decrease with increasing pair transverse momentum. These trends for K$^\pm$K$^\pm$ are similar to the ones observed earlier in identical charged-pion and K$_{\rm s}^{0}$K$_{\rm s}^{0}$ correlations in Pb--Pb collisions at various energies and in $\pi^\pm \pi^\pm$ correlations in p--Pb collisions at $\sqrt{s{_{\rm NN}}} = 5.02$~TeV. At comparable multiplicity, the source sizes measured in p--Pb collisions agree within uncertainties with those observed in pp collisions, and there is an indication that they are smaller than those observed in Pb--Pb collisions. The obtained results are also compared with predictions from the hadronic interaction model EPOS~3, which tends to underestimate the source size for the most central collisions and agrees with the data for semicentral and peripheral events. Furthermore, the time of maximal emission for kaons is extracted. It turns out to be comparable with the value obtained in highly peripheral Pb--Pb collisions at the same energy, indicating that the kaon emission evolution is similar to that in p--Pb collisions.
\end{abstract}
\end{titlepage}
\setcounter{page}{2}

\section{Introduction}
During its operation, the Large Hadron Collider (LHC)~\cite{Evans:2008zzb} delivered pp, p--Pb, and Pb--Pb collisions at various energies, which have been used to study the properties of matter created in such environments. Signatures of quark--gluon plasma (QGP) formation were observed in Pb--Pb collisions, such as a strong suppression of high-momentum particle production (jet quenching)~\cite{CMS:2021vui,ALICE:2019qyj,ATLAS:2022zbu}, a suppression of the K$^*$(892)$^0$/K yield ratio~\cite{ALICE:2017ban,ALICE:2019xyr}, strangeness enhancement~\cite{NA49:2002fxd,ALICE:2013xmt}, Debye screening effect~\cite{Faccioli:2018hjk,CMS:2012gvv,ALICE:2020wwx,ALICE:2014ict}, or collective flow effects 
\cite{CMS:2022bmk,ALICE:2022zks}, which are well described by hydrodynamic models~\cite{Gale:2013da,Heinz:2013th,Romatschke:2017ejr}. As shown by the comparison of the Pb--Pb results to reference pp measurements, these effects could not be described by a simple incoherent superposition of nucleon--nucleon interactions and could be interpreted as final-state phenomena, characteristic to QGP creation in heavy-ion collisions~\cite{BRAHMS:2004adc,STAR:2005gfr,CMS:2012aa,ATLAS:2012tjt,ALICE:2022wpn}. Since an extended QGP phase is not expected to form in proton--nucleus collision systems, the comparison of the p--Pb results to the Pb--Pb ones might help select observables that could be used to understand the mechanism and conditions of QGP formation in high-energy collisions.

The femtoscopy technique~\cite{Goldhaber:1960sf,Kopylov:1974th,Lednicky:2005af,Lisa:2005dd}, which studies particle correlations in momentum space, is an effective tool for extracting the size of the particle-emitting region at freeze-out~\cite{Heinz:2007in} (conventionally called the ``radius'' parameter) and the decoupling time of matter created in relativistic collisions. Strictly speaking, by using femtoscopy one can measure not the whole volume of the particle emitting source but the size of the region of an expanding source from which the particles with similar momenta are emitted~\cite{Akkelin:1995gh,Bowler:1985eny}, the so-called ``homogeneity volume''. This volume is smaller than the total volume of the system and decreases with increasing pair transverse momentum $k_{\rm T}=|{\bf p}_{\rm T,1}+{\bf p}_{\rm T,2}|/2$ or transverse mass $m_{\rm T}=\sqrt{m^2+k_{\rm T}^2}$ (for particles with identical masses $m$ and transverse momenta ${\bf p}_{\rm T,1}$ and ${\bf p}_{\rm T,2}$). Thus, the femtoscopic correlations can be used to measure the spatial and temporal information of the particle emitting source allowing for extraction of the dynamic information on the emission process.

In the one-dimensional (1D) femtoscopic analysis, the source, in the pair rest frame, 
is characterized by one size parameter $R_{\rm inv}$~\cite{Lisa:2005dd}. Several 1D analyses were performed at the LHC for correlations of various particle species in pp~\cite{CMS:2011nlc,ALICE:2012yyu,ALICE:2012aai,CMS:2019fur,ALICE:2021ovd}, p--Pb~\cite{ALICE:2019kno,LHCb:2023dcc}, and Pb--Pb collisions~\cite{ALICE:2015hvw,ALICE:2014xrc,ALICE:2022mxo}. Since the 1D analysis can be performed even with a small amount of collected experimental data, it can be done in almost any experiment and for various types of particles. However, the information obtained in this kind of analysis is limited by extracting a generalized size, and no information on the shape of the source can be obtained. 
In the more general case, the shape of the source can be retrieved in the three-dimensional (3D) analysis where its size is determined independently for three directions: ``long'' ($R_{\rm long}$) along the beam direction ($z$ axis of the reference frame), ``out'' ($R_{\rm out}$) along the pair transverse momentum, and ``side'' ($R_{\rm side}$), perpendicular to the other two. For high-energy collisions, analyses are usually performed in the longitudinally comoving system (LCMS), a rest frame moving along the beam direction such that the pair total momentum projection $P_z$ vanishes, i.e. $P_z=P_{1,z} + P_{2,z} = 0$~\cite{Lisa:2005dd}. The 3D analysis requires more experimental data and, therefore, is often performed for the most abundantly produced particles, i.e. charged pions~\cite{ALICE:2011kmy,ALICE:2015hav,ATLAS:2017shk}. For kaons, the second most copiously produced particles, all 3D femtoscopic analyses that have been done by now are for heavy-ion collisions~\cite{E802:1997muu,PHENIX:2009ilf,ALICE:2017iga}.
For proton--nucleus collision systems, for example p--Pb collisions, only a few 1D femtoscopic analyses have been performed so far~\cite{Beker:1994qs,ALICE:2019kno}. The study of femtoscopic correlations in such systems is particularly interesting because it provides a bridge between small (pp) and large (A--A) collision systems, and may lead to additional constraints on model scenarios that successfully describe pp and A--A collisions. While the A--A results can be interpreted within the hydrodynamic framework~\cite{Lisa:2005dd,Soltz:2012rk,Karpenko:2012yf}, and the pp data are explained by a hydrodynamic expansion~\cite{Hirono:2014dda} taking into account additional effects related to the uncertainty principle~\cite{Shapoval:2013jca}, there is no common interpretation of the results obtained in collisions where one of the colliding projectiles is a particle. Some studies~\cite{PHENIX:2014dmi,Bozek:2014fxa,LHCb:2023dcc} suggest that a hydrodynamic evolution similar to that for A--A collisions may be present in such systems. However, as shown in Refs.~\cite{ALICE:2014xrc,ALICE:2015hav,ALICE:2019kno}, the radii for the p--Pb system, at a given track multiplicity, were found to be closer in magnitude to those in pp, while there was a visible gap separating the radii in p--Pb and Pb--Pb collisions. This may demonstrate the influence of different initial conditions on the final-state, or indicate significant collective expansion already in peripheral Pb--Pb collisions.

In this work, 1D and 3D femtoscopic correlations of two identical charged kaons K$^\pm$K$^\pm$ are studied in p--Pb collisions at $\sqrt{s_{\rm NN}} = 5.02$~TeV. Identical-kaon femtoscopy gives information about the collision region that is complementary to that obtained with identical-pion femtoscopy. It probes the hotter region of the particle emitting source, where strange quarks are produced, and extends the momentum range over which the femtoscopy analysis can be applied. The kaon analysis also offers a cleaner signal compared to pions, as it is less affected by resonance decays. The kaon femtoscopic radii as a function of the pair transverse momentum, which are reported in this article, provide additional constraints on the validity of approaches used to describe matter created in ultrarelativistic collisions. Comparison of femtoscopic characteristics in different collision systems, such as pp, p--Pb, and Pb--Pb, and with EPOS~3.111 model~\cite{Werner:2013tya} calculations allows investigating the dynamics of high-energy collision evolution and also helps tune and constrain the model parameters. Fitting the $m_{\rm T}$ dependence of $R^2_{\rm long}$ in the LCMS~\cite{Csorgo:1995bi,Danielewicz:2006hi} with the function suggested in Refs.~\cite{Shapoval:2020nec,Sinyukov:2015kga}, the duration of the collision evolution characterized by the time of maximal emission can be estimated. It was shown in the analysis of identical charged kaon correlations in Pb--Pb collisions at $\sqrt{s_{\rm NN}}=5.02$ TeV~\cite{GRomanenko_AN} and its comparison with the integrated hydrokinetic model (iHKM) calculations~\cite{Shapoval:2021fqg} that the system created in peripheral collisions evolves faster than in central ones. In the presented analysis, the time of maximal emission for kaons $\tau_{\rm K}$ was extracted from the experimental data for an asymmetric collision system, namely p--Pb, to understand the particle emission dynamics by comparing these results with those obtained in Pb--Pb collisions.

The article is organized as follows: Sec.~\ref{sec:data} describes the ALICE apparatus, the analyzed data sample, and the event- and track-selection criteria. The details of the femtoscopic correlation function analysis presented in this work are given in Sec.~\ref{CFan}. The associated systematic uncertainties are discussed in Sec.~\ref{SystematicUncertainties}. The obtained results are shown and compared with model predictions in Sec.~\ref{sec:results} and summarized in Sec.~\ref{sec:summary}.

\section{Data analysis}
\label{sec:data}
The data sample studied in this work was collected in 2016 by ALICE from p--Pb collisions at a center-of-mass energy per nucleon--nucleon pair of $\sqrt{s_{\mathrm{NN}}}=5.02$ TeV during the LHC Run 2 period.
The nucleon--nucleon center-of-mass system is shifted with respect to the ALICE laboratory system by 0.465 units of rapidity in the direction of the proton beam~\cite{ALICE:2004fvi}. Throughout this paper $\eta$ represents the pseudorapidity measured in the laboratory frame.

In the LHC Run 2 period, the ALICE apparatus consisted of a central barrel, covering the pseudorapidity region $|\eta| < 0.9$, a muon spectrometer with $-4.0 < \eta < -2.5$ coverage, and forward- and backward-pseudorapidity detectors employed for triggering, background rejection, and event characterization. An overview of the detector and its performance are presented in Refs.~\cite{ALICE:2008ngc,ALICE:2014sbx,ALICE:2022wpn}.

Approximately $250\times10^6$ minimum-bias events were analyzed. They were obtained using the minimum-bias trigger provided by the V0 detector~\cite{ALICE:2004ftm}, which is composed of two scintillator arrays placed on either side of the interaction point along the beam axis with pseudorapidity coverage $2.8 < \eta < 5.1$ (V0A, located on the Pb remnant side) and $-3.7 < \eta < -1.7$ (V0C). The minimum-bias trigger
requires a signal in both V0 detectors within a time window
that is consistent with the collision occurring at the center of
the ALICE detector. The V0 detector was also utilized for determining the multiplicity of the analyzed events~\cite{ALICE:2013wgn} using the measured energy deposition. The multiplicity classes were defined in terms of percentile intervals of the measured V0A amplitude~\cite{ALICE:2014xsp}. The central barrel detectors used in the analysis are the Inner Tracking System (ITS)~\cite{ALICE:1999cls,ALICE:2010tia}, the Time Projection Chamber (TPC)~\cite{ALICE:2000jwd,Alme:2010ke}, and the Time-Of-Flight detector (TOF)~\cite{ALICE:2000xcm,ALICE:2002imy,Akindinov:2013tea}. They are embedded in a large solenoidal magnet that provides a magnetic field of 0.5~T parallel to the beams axis. The ITS consists of six layers of silicon detectors and was used to reconstruct the primary vertex and to track charged particles. The TPC is the main tracking detector of the central barrel. It is a gaseous detector placed co-axially with the beam axis and radially next to the ITS. It is divided by a central electrode into two halves. The end caps on either side are composed of 18 sectors (covering the full azimuthal angle) with 159 pad rows placed radially in each sector. A track signal in the TPC consists of space points (clusters), each of which is reconstructed in one of the pad rows. It also enables charged-particle identification via the measurement of the particle specific energy loss (d$E$/d$x$) in the detector gas. The TPC covers an acceptance of $|\eta| < 0.9$ for tracks that reach the outer radius of the detector and $|\eta| < 1.5$ for shorter tracks. The TOF is a gaseous detector, which uses multigap resistive plate chambers~\cite{CerronZeballos:1995iy} as its basic detecting element. The primary-vertex position was determined with tracks reconstructed in the ITS and TPC, as described in Ref.~\cite{ALICE:2012aqc}. In order to obtain a uniform acceptance of the detectors, only events with a reconstructed primary vertex within $\pm10$~cm from the center of the detector along the beam direction $z$ were included in the analysis. The parameters of each track are determined by performing a Kalman fit to a set of clusters with an additional constraint for the track to pass through the primary vertex. The fit was required to deliver $\chi^2/{\rm NDF}$ less than~2. The transverse momentum of each track was determined from its curvature in the magnetic field. The particle identification capabilities of the TPC are supplemented with the TOF detector, which provides a measurement of the time of flight of charged particles.

\subsection{Event classification}
The 1D analysis was conducted in two different variants. The first was performed in six multiplicity classes~\cite{ALICE:2013wgn,ALICE:2014xsp}: 0--5\%, 5--10\%, 10--20\%, 20--40\%, 40--60\%, and 60--80\% and two pair transverse momentum $k_{\rm T}$ ranges: (0.2--0.5) and (0.5--1.0)~GeV$/c$ for a finer comparison with the results from pp~\cite{ALICE:2012aai} and Pb--Pb~\cite{ALICE:2015hvw,GRomanenko_AN} collision systems and also for crosscheck with the results of the previous p--Pb analysis~\cite{ALICE:2019kno}. The second was done in three multiplicity classes: 0--20\%, 20--40\%, and 40--90\% and in eight $k_{\rm T}$ ranges: (0.2--0.3), (0.3--0.4), (0.4--0.5), (0.5--0.6), (0.6--0.7), (0.7--0.8), (0.8--1.0), and (1.0--1.3)~GeV$/c$ to better study the trend with $k_{\rm T}$ and to have a more detailed comparison with the EPOS~3 calculations. The 3D analysis was performed in three multiplicity classes: 0--20\%, 20--40\%, and 40--90\% and two $k_{\rm T}$ ranges: (0.2--0.5) and (0.5--1.0)~GeV$/c$.

Tables~\ref{tab:KKmultiplicities1} and~\ref{tab:KKmultiplicities2} show the mean charged-particle multiplicity densities $\langle {\rm d}N_{\rm ch}/{\rm d}\eta\rangle$ averaged over $|\eta|<0.5$ using the method presented in Ref.~\cite{ALICE:2013wgn} for the analyses in six and three multiplicity classes, respectively. The $\langle {\rm d}N_{\rm ch}/{\rm d}\eta\rangle$ values were not corrected for trigger and vertex-reconstruction inefficiency, which is about 4\% for non-single diffractive events~\cite{ALICE:2012xs}.
\begin{table}[ht]
\centering
\caption{Event multiplicity classes and their corresponding $\langle {\rm d}N_{\rm ch}/{\rm d}\eta\rangle$~\cite{ALICE:2013wgn} for the analysis in six multiplicity intervals as obtained using the method presented in Ref.~\cite{ALICE:2012xs}. The given uncertainties are systematic only, the statistical uncertainties are negligible.}
\begin{tabular}{cc}
  \hline\hline
    Event class  &  $\langle {\rm d}N_{\rm ch}/{\rm d}\eta\rangle$, $|\eta|<0.5$\\ \hline
    0--5\%       &  45.0$\pm$1.0\\
    5--10\%      &  36.2$\pm$0.8\\
    10--20\%      &  30.5$\pm$0.7\\
    20--40\%      &  23.2$\pm$0.5\\
    40--60\%      &  16.1$\pm$0.4\\
    60--80\%      &  9.8$\pm$0.2\\
    \hline\hline
\end{tabular}
\label{tab:KKmultiplicities1}
\end{table}
\begin{table}[ht]
\centering
\caption{Event multiplicity classes and their corresponding $\langle {\rm d}N_{\rm ch}/{\rm d}\eta\rangle$~\cite{ALICE:2013wgn} for the analysis in three multiplicity intervals as obtained using the method presented in Ref.~\cite{ALICE:2012xs}. The given uncertainties are systematic only, the statistical uncertainties are negligible.}
\begin{tabular}{cc}
  \hline\hline
    Event class  &  $\langle {\rm d}N_{\rm ch}/{\rm d}\eta\rangle$, $|\eta|<0.5$\\ \hline
    0--20\%       &  35.6$\pm$0.8\\
    20--40\%      &  23.2$\pm$0.5\\
    40--90\%      &  12.9$\pm$0.3\\
    \hline\hline
\end{tabular}
\label{tab:KKmultiplicities2}
\end{table}
\subsection{Particle and pair selection} \label{ChargedKaonSelection}

The identification of kaons was conducted employing both the TPC and the TOF detectors by applying a strict selection on the deviation $n_\sigma$ between the measured quantities (d$E$/d$x$ and time-of-flight) and the expected values for a kaon, normalized by the detector resolution $\sigma$. The $n_\sigma$ thresholds were chosen so as to remove possible contamination from electrons, pions, and protons to the kaon sample. The kaon candidates were selected within a transverse momentum range of $0.14<p_{\rm T}<1.50$~GeV$/c$ and a pseudorapidity range of $|\eta| < 0.8$, to avoid regions of the detector with limited acceptance. To significantly improve the amount of primary kaons with respect to secondary particles coming from weak decays and particle--detector interactions, a selection criterion on the distance of closest approach (DCA) to the primary vertex was applied, both in the transverse plane (${\rm DCA}_{xy} < 0.135$~cm) and along the direction of the beam (${\rm DCA}_z < 0.13$~cm). Tracks reconstructed in the TPC were selected based on the criteria listed in Table~\ref{tab:KKcuts}, which are similar to those used in the analysis described in Ref.~\cite{ALICE:2019kno}. To ensure a good momentum resolution, each track was required to be composed of at least 80 out of the maximum possible 159 TPC clusters. Charged-particle tracks with momentum $p < 0.5$~GeV$/c$ were identified as kaons if they satisfied the requirement $|n_{\sigma,{\rm TPC}}|<2$. Tracks with $p > 0.5$~GeV$/c$ were required to match to a signal in the TOF, and satisfy $|n_{\sigma,{\rm TPC}}|<3$ as well as the following momentum-dependent $n_{\sigma,{\rm TOF}}$ selection: $|n_{\sigma,{\rm TOF}}|<2$ for $0.5<p<0.8$~GeV$/c$, $|n_{\sigma,{\rm TOF}}|<1.5$ for $0.8 < p< 1.0$~GeV$/c$, and $|n_{\sigma,{\rm TOF}}|<1$ for $1.0 < p< 1.5$~GeV$/c$.

\begin{table}
\centering
\caption{Charged kaon selection criteria.}
\begin{tabular}{ll}
  \hline\hline
    Selection criterium & Value \\ \hline
    $p_{\rm T}$  & $0.14<p_{\rm T}<1.5$~GeV$/c$ \\ \hline
    $|\eta|$ & $< 0.8$ \\ \hline
    $\rm DCA_{\it xy}$ & $< 0.135$~cm \\ \hline
    $\rm DCA_{\it z}$ & $< 0.13$~cm \\ \hline
    \multirow{2}{*}{$|n_{\sigma,{\rm TPC}}|$} & $< 2$ (for $ 0.14 < p < 0.50$~GeV$/c$) \\
    & $< 3$ (for $p > 0.5$~GeV$/c$) \\ \hline
     & $< 2$ (for $0.5 < p < 0.8$~GeV$/c$) \\
    $|n_{\sigma,{\rm TOF}}|$ & $< 1.5$ (for $0.8 < p < 1.0$~GeV$/c$) \\
     & $< 1.0$ (for $1.0 < p < 1.5$~GeV$/c$) \\ \hline
    Number of track points in TPC  & $\geq$~80 \\ \hline\hline
\end{tabular}
\label{tab:KKcuts}
\end{table}

As discussed in detail in Ref.~\cite{ALICE:2011kmy}, the femtoscopic correlation functions of two identical particles are sensitive to two-track reconstruction effects when the particles are close in momentum and have close trajectories. Two kinds of two-track effects were investigated in this work. Track ``splitting'' occurs when one track is mistakenly reconstructed as two. Track ``merging'' corresponds to the reconstruction of two different tracks as one when they cannot be distinguished if their trajectories stay close to each other through a significant part of the TPC volume. To suppress such cases, each cluster in the TPC was flagged as ``shared'' if it was used in the reconstruction of more than one track, and pairs that shared $>5\%$ of clusters in the TPC were removed from the sample. In addition, the selected pairs of kaons were required to be separated on average in the TPC detector by at least 3~cm.

\subsection{Purity}
\begin{figure}[h]
\centering
\subfigure{\label{fig:single_purity}\includegraphics[width=0.49\linewidth]{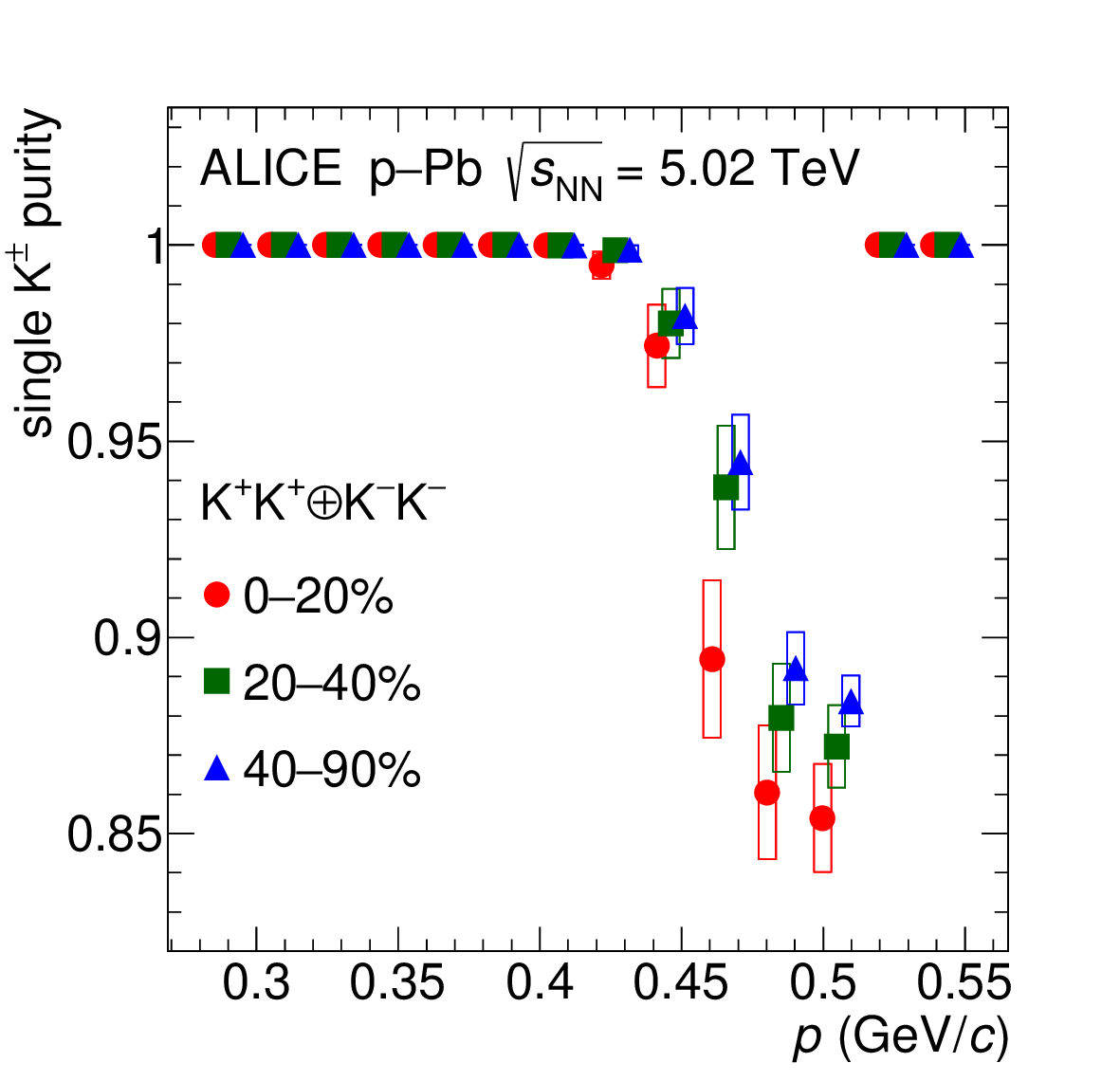}}
\subfigure{\label{fig:pair_purity}\includegraphics[width=0.49\linewidth]{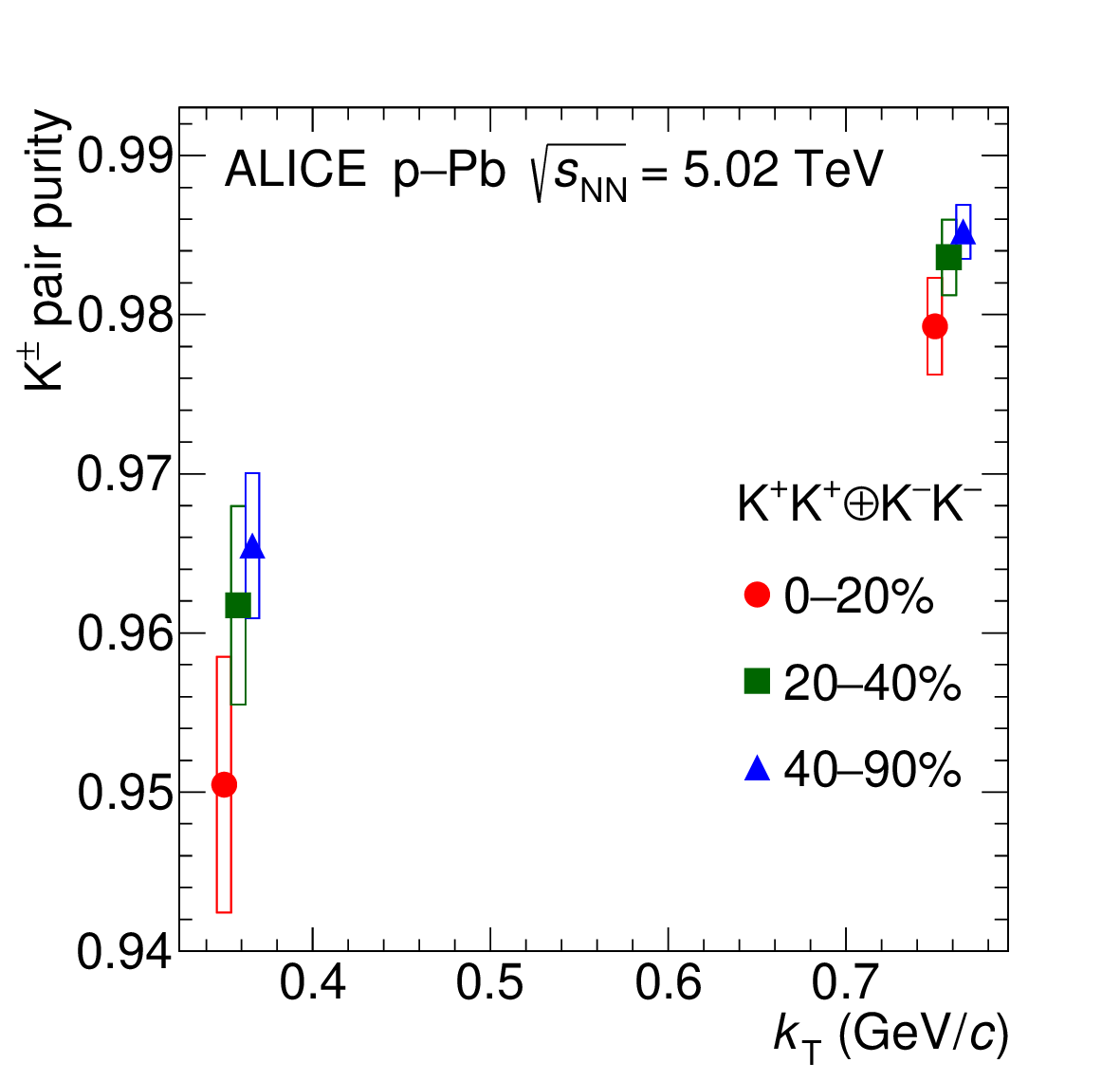}}
\caption{(Color online) Single (left) and pair (right) K$^\pm$ purities for different event multiplicity classes. The systematic uncertainties associated with the purity estimation are shown as boxes. Statistical uncertainties are negligible. The momentum $p$ and $k_{\rm T}$ values for lower multiplicity classes (blue and green symbols) are slightly shifted for clarity.}\label{fig:purity}
\end{figure}

The kaon purity for $p<0.5$~GeV/$c$ was estimated as in Ref.~\cite{ALICE:2019kno} by parametrizing the TPC d$E$/d$x$ distribution of the experimental data in momentum slices and computing the fraction of particle species that could mistakenly contribute to the kaon signal. The purity of the kaon sample was defined as the fraction of accepted kaon tracks that correspond to true kaons.
The dominant contamination for charged kaons comes from e$^\pm$ in the momentum range $0.4<p<0.5$~GeV$/c$. The parameters of the function fitting the TPC distribution in momentum slices depend on the fit interval and are the source of the systematic uncertainty associated with the estimated single kaon purity. The purity for $p>0.5$~GeV$/c$, where the TOF information was employed, was studied with DPMJET~\cite{Roesler:2000he} simulations using GEANT~3~\cite{Brun:1994aa} to model particle transport through the detector. Based on the results of this study, the $n_{\sigma,{\rm TOF}}$ thresholds were chosen to provide a single kaon purity greater than 99\%. The momentum and multiplicity dependence of the single kaon purity in the region of maximal contamination is shown in Fig.~\ref{fig:purity} (left panel). The pair purity was calculated as the product of two single-particle purities for pairs with relative momentum $q_{\rm inv}<0.25$~GeV$/c$ ($q_{\rm inv}$ is defined below in Sec.~\ref{CFan}), where the momenta were taken from the experimentally determined distribution. The obtained K$^\pm$ pair purity as a function of $k_{\rm T}$ is shown in Fig.~\ref{fig:purity} (right panel). The pair purity is higher than the single-kaon one due to the effect of averaging over low- and high-momentum ranges in the full single-kaon momentum interval $p$ (see Ref.~\cite{ALICE:2019kno} for details).

\section{\label{CFan}Measurement and parametrization of the correlation function}
The correlation function (CF) of two particles with momenta ${\bf p}_1$ and ${\bf p}_2$ is measured as a function of the momentum difference of the pair ${\bf q} = {\bf p}_1-{\bf p}_2$ and can be expressed as the ratio
\begin{equation}
C({\bf q})=\frac{A({\bf q})}{B({\bf q})}\label{eq:CFdef}
\end{equation}
of the two-particle distribution in a given event $A({\bf q})$ to the reference distribution $B({\bf q})$~\cite{Kopylov:1974th}. The distribution $A({\bf q})$ was constructed from the events having at least two identical charged kaons.
The reference distribution was formed by mixing the events containing at least one charged kaon, where each event was mixed with five other events, which had similar $z$ position (within 2~cm) of the primary vertex and similar multiplicity (within $1/4$ of the width of the given multiplicity class)~\cite{Lisa:2005dd}. The correlation function $C$ is normalized to unity such that $C\rightarrow1$ in the absence of a correlation signal.

The 1D CF was calculated in the pair rest frame (where ${\bf p}_1+{\bf p}_2=0$) as a function of the invariant relative momentum of the pair $q_\mathrm{inv}= \sqrt{|{\bf q}|^{2} - q_{0}^{2}} $, where $q_0=E_1-E_2$ and ${\bf q}={\bf p}_1-{\bf p}_2$ are determined by the energies $E_1$, $E_2$ and momenta ${\bf p}_1$, ${\bf p}_2$ of the particles, respectively. In the 3D femtoscopic analysis, the momentum difference ${\bf q}$ was evaluated in the LCMS.

The correlation functions were corrected for purity according to~\cite{ALICE:2015hvw,ALICE:2017iga,ALICE:2019kno} as
\begin{equation}
C_{\rm corrected} = (C_{\rm raw}-1+P)/P,
\label{eq:PurCorr}
\end{equation}
where the pair purity $P$ depends on $k_{\rm T}$ and multiplicity, as shown as an example for the case of the 3D analysis in Fig.~\ref{fig:purity} (right panel).

The effect of finite track momentum resolution in the 1D analysis was accounted for through the use of a response matrix generated with DPMJET simulations and was applied on-the-fly during the fitting process (see Sec.~\ref{1DAnalisys} below) as described in Ref.~\cite{ALICE:2020wvi}. In the 3D case, the correlation function was corrected for a momentum resolution factor evaluated with DPMJET simulations as the ratio of the CF as a function of the reconstructed relative momentum of the pair to the CF as a function of the true pair relative momentum. The details of the procedure can be found in Ref.~\cite{ALICE:2017iga}.

To extract information on the particle emitting source from the measured correlation function, the CFs are parametrized by various formulas depending on the origin of the correlations between the considered particles. For the K$^\pm$K$^\pm$ case, they are quantum statistics and the Coulomb interaction. The effects in the CF due to strong final-state interactions between identical charged kaons are negligible given the shallow interaction~\cite{Beane:2007uh}.

\subsection{Fitting 1D correlation function}\label{1DAnalisys}
Considering a Gaussian distribution for a spherical particle source in the pair rest frame (PRF), the fit of the 1D CF was performed using the Bowler--Sinyukov formula~\cite{Bowler:1985eny,Sinyukov:1998fc}
\begin{equation}
C(q_\mathrm{inv})=N\left[1 -\lambda +\lambda K(r,q_\mathrm{inv})\left( 1+
\exp{\left(-R_\mathrm{inv}^{2} q^{2}_\mathrm{inv}\right)}\right)\right]\,D(q_\mathrm{inv}),
\label{eq:BS_CF}
\end{equation}
where $D(q_\mathrm{inv})$ is a function used to parametrize all non-femtoscopic effects such as resonance decays or mini-jets~\cite{ALICE:2011kmy,ALICE:2012aai}, and $N$ is a normalization coefficient. The factor $K(r,q_\mathrm{inv})$ describes the Coulomb interaction with a radius $r$ evaluated according to Refs.~\cite{Sinyukov:1998fc,ALICE:2013uhj} as
\begin{eqnarray}
K(r,q_\mathrm{inv})=\frac{C_{\rm QS+Coulomb}}{C_{\rm QS}},\label{Coulomb}
\end{eqnarray}
where the theoretical correlation function $C_{\rm QS}$ takes into account quantum statistics (QS) only and $C_{\rm QS+Coulomb}$ includes quantum statistical and Coulomb final-state interaction effects evaluated using the ${\rm Lednick\acute{y}}$ model~\cite{Lednicky:2005af}.
The radius $r$ determines the range of the Coulomb interaction used to calculate the $C_{\rm QS+Coulomb}$ value. It was set to $r=1.5$~fm, which is on average close to the radius values extracted in the presented analysis. Its uncertainty has systematic effects on the final results (as discussed in Sec.~\ref{SystematicUncertainties} below).

The parameters $R_\mathrm{inv}$ and $\lambda$ describe the size of the source and the correlation strength, respectively.
They were extracted from the fit of Eq.~(\ref{eq:BS_CF}) to the data in the range of the correlation signal $0<q_\mathrm{inv}<0.5$~GeV$/c$ handling the baseline $D(q_\mathrm{inv})$ as described in Ref.~\cite{ALICE:2019kno}. The non-femtoscopic effects were estimated with the EPOS~3 model~\cite{Werner:2013tya} without QS nor Coulomb interaction effects included. The CF was normalized to unity outside the femtoscopic effect region, i.e. in the $0.25<q_\mathrm{inv}<0.50$~GeV$/c$ range. The example 1D correlation functions can be found in Ref.~\cite{ALICE:2019kno}. These CFs show all the features inherent in femtoscopic correlations of identical charged kaons in p--Pb collisions at $\sqrt{s_{\rm NN}}=5.02$~TeV, which are similar independently of the used multiplicity classes and $k_{\rm T}$ ranges.

\subsection{Fitting 3D correlation function}
\begin{figure}[ht!]
\centering
{\includegraphics[width=0.85\linewidth]{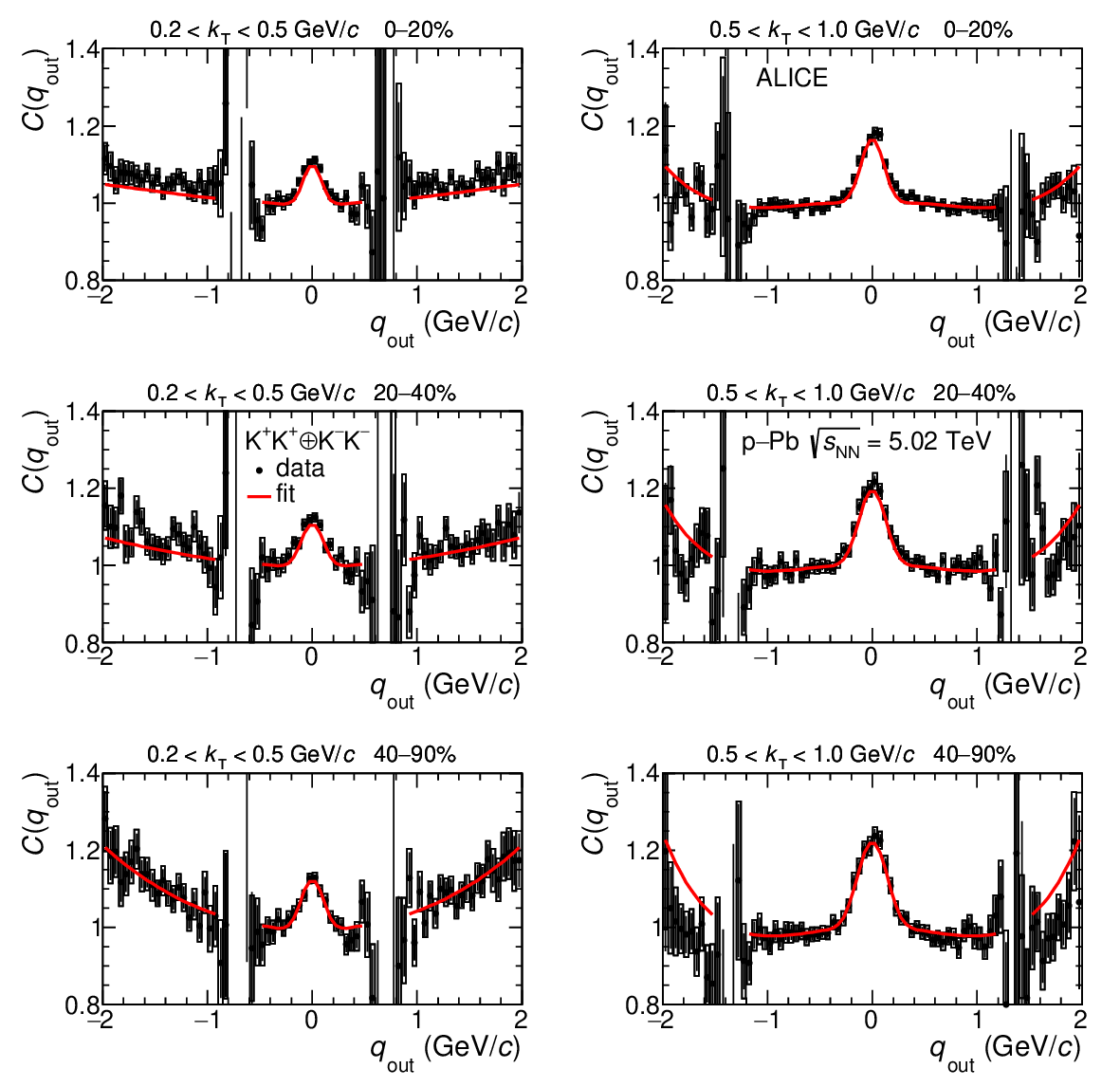}}
\caption{(Color online) Projection of the 3D CF in the ${\rm out}$ direction (black full circles) in three multiplicity classes and two $k_{\rm T}$ ranges fitted (red solid lines) with Eq.~(\ref{eq:BS_3DCF}). Statistical (systematic) uncertainties are shown by bars (boxes). The data are integrated over the range of $|q_i| < 0.15$~GeV$/c$ in the non-projected coordinates. Vertical lines not belonging to any marker around $|q_{\rm out}|=0.6$~GeV$/c$ are very-large error bars of data points that are out of the chosen $Y$ axis range.
} \label{fig:CF3D_poly_out}
\end{figure}
%
\begin{figure}[ht!]
\centering
{\includegraphics[width=0.85\linewidth]{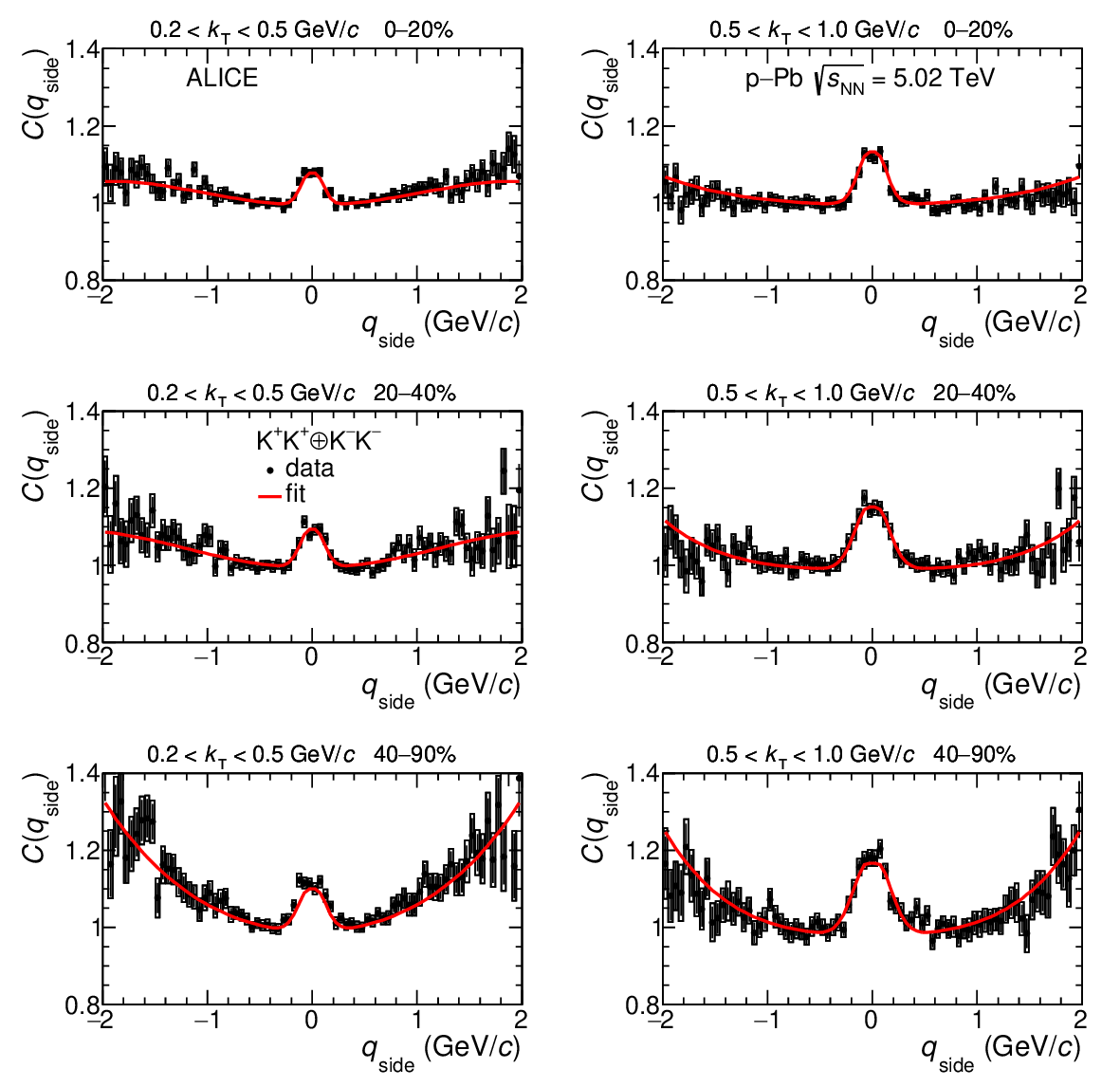}}
\caption{(Color online) Projection of the 3D CF in the ${\rm side}$ direction (black full circles) in three multiplicity classes and two $k_{\rm T}$ ranges fitted (red solid lines) with Eq.~(\ref{eq:BS_3DCF}). Statistical (systematic) uncertainties are shown by bars (boxes). The data are integrated over the range of $|q_i| < 0.15$~GeV$/c$ in the non-projected coordinates.
} \label{fig:CF3D_poly_side}
\end{figure}
\begin{figure}[ht!]
\centering
{\includegraphics[width=0.85\linewidth]{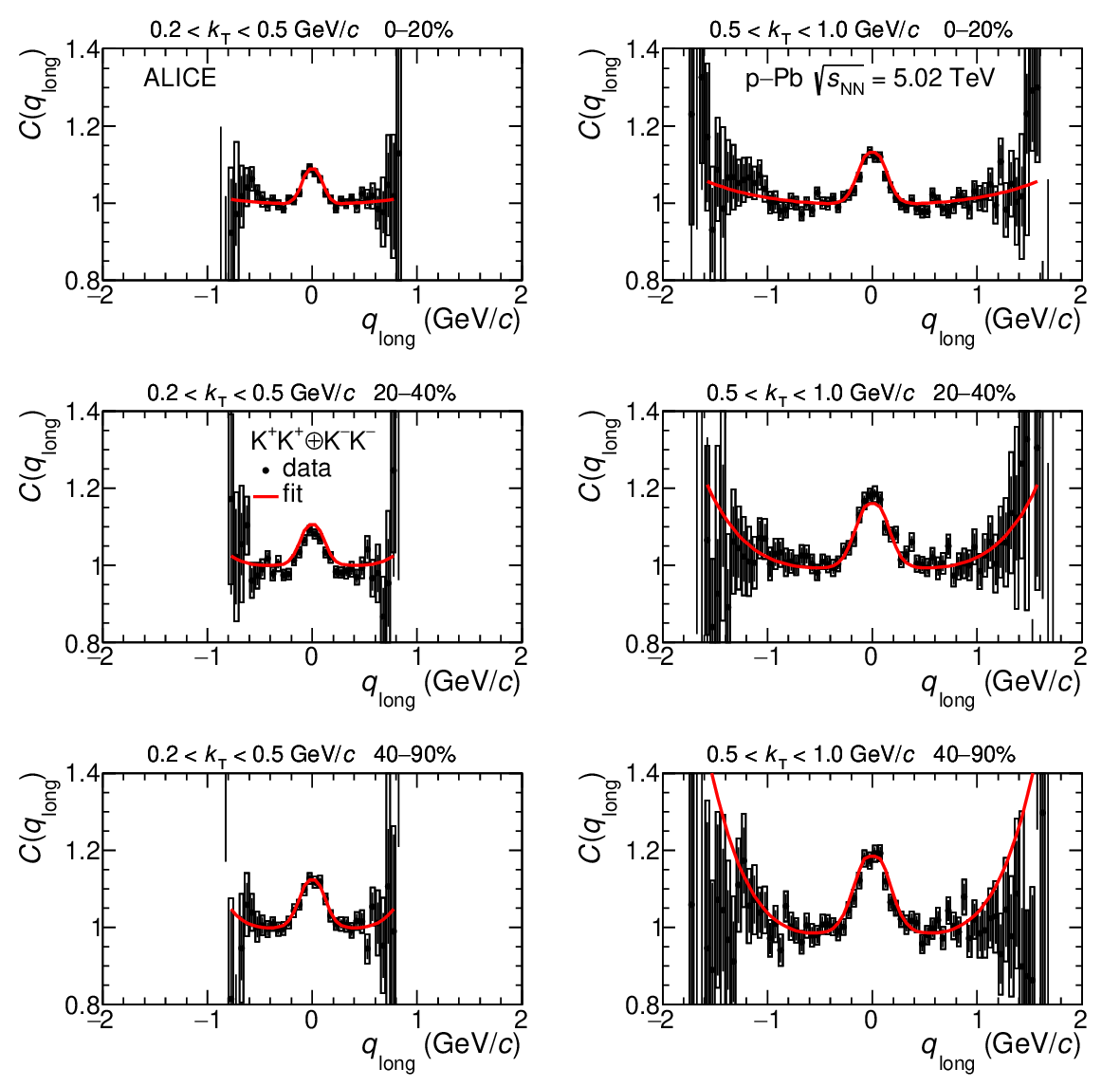}}
\caption{(Color online) Projection of the 3D CF in the ${\rm long}$ direction (black full circles) in three multiplicity classes and two $k_{\rm T}$ ranges fitted (red solid lines) with Eq.~(\ref{eq:BS_3DCF}). Statistical (systematic) uncertainties are shown by bars (boxes). The data are integrated over the range of $|q_i| < 0.15$~GeV$/c$ in the non-projected coordinates.
Vertical lines not belonging to any marker at $|q_{\rm long}|\gtrapprox1.0$~GeV$/c$ are very-large error bars of data points that are out of the chosen $Y$ axis range.
} \label{fig:CF3D_poly_long}
\end{figure}
To extract the information about the 3D size and shape of the particle emitting source, the 3D CF was fitted with the Bowler--Sinyukov formula (for a Gaussian-like source)~\cite{Bowler:1991vx,Sinyukov:1998fc}
\begin{eqnarray}
C({\bf q})=N\left[1-\lambda+\lambda K(q_{\rm inv},R_{\rm inv})\left(1+\exp\left(-R_{\rm out}^2q_{\rm out}^2-R_{\rm side}^2q_{\rm side}^2-R_{\rm long}^2q_{\rm long}^2\right)\right)\right]D\left({\bf q}\right),\label{eq:BS_3DCF}
\end{eqnarray}
where $q_{\rm out}$, $q_{\rm side}$, and $q_{\rm long}$ are projections of the pair relative momentum ${\bf q}$ in the LCMS, and $R_{\rm out}$, $R_{\rm side}$, and $R_{\rm long}$ are the corresponding femtoscopic source radii, with $R_{\rm out}$ related to the geometrical size of the emitting source and particle emission duration, $R_{\rm side}$ to the geometrical size, and $R_{\rm long}$ to the time of maximal emission. The parameters $N$, $\lambda$, $K(q_{\rm inv}, R_{\rm inv})$, and $D\left(q_{\rm out},q_{\rm side},q_{\rm long}\right)$ have the same meaning as in the 1D analysis in Eq.~(\ref{eq:BS_CF}). The normalization range was $0.3 < q_i < 0.4$~GeV$/c$ where $i$ = out, side, long. The function $K(q_{\rm inv},R_{\rm inv})$ is the Coulomb part of the two-kaon wave function integrated over the spherical Gaussian source with a fixed radius $R_{\rm inv}$ calculated in the PRF. The average $q_{\rm inv}$ in the PRF for the given ``out-side-long'' interval was determined during the $C({\bf q})$ construction. The value of this radius was chosen to be 1.5~fm like in the 1D analysis (see Sec.~\ref{1DAnalisys}).

The CF was fitted with Eq.~(\ref{eq:BS_3DCF}) in the $|q_i| < 1.5$~GeV$/c$ range. The baseline was parametrized with a fourth-order polynomial
\begin{eqnarray}
D\left({\bf q}\right) = 1+a_{\rm out}q_{\rm out}^2+a_{\rm side}q_{\rm side}^2+a_{\rm long}q_{\rm long}^2+b_{\rm out}q_{\rm out}^4+b_{\rm side}q_{\rm side}^4+b_{\rm long}q_{\rm long}^4,\label{eq:D_poly}
\end{eqnarray}
which reasonably describes the CF outside the femtoscopic peak region. The results are shown in Figs.~\ref{fig:CF3D_poly_out}--\ref{fig:CF3D_poly_long}. As can be seen from the figures, the fit reproduces well the shape of the correlation peak and also captures non-femtoscopic behavior of all projections of the 3D correlation function. The areas with no data points in the $q_{\rm out}$ projection around $\pm0.6$~GeV$/c$ and in the $q_{\rm long}$ projection below $-0.8$ (and above 0.8)~GeV$/c$ originate from gaps in the acceptance due to the kinematical selections used in the analysis and the limited pseudorapidity acceptance of the detector, respectively~\cite{ALICE:2011kmy}.
\section{Systematic uncertainties}\label{SystematicUncertainties}

The systematic uncertainties were estimated by varying the criteria used for the selection of the events, tracks, and pairs and the fit parameters. The criteria used for particle identification and pair selection were varied by up to $\pm$20$\%$.
The influence of the fit range was investigated by shifting the $q_{\rm inv}$ ($q_{\rm out, side, long}$) upper (left and right) limit(s) by $\pm$20$\%$.
The range of the primary-vertex position from the center of the detector along the beam direction $z$ for reconstructed events was varied by $\pm$10$\%$.

There is also an uncertainty associated with the choice of the radius of the Coulomb interaction. By default, it was set to $r=1.5$~fm as a result of averaging the values of the femtoscopic radii that were extracted from the data for the considered multiplicity classes and $k_{\rm T}$ ranges and varied by $\pm0.5$~fm.

The fitting procedure requires knowledge of the non-femtoscopic background shape and magnitude. In this work, the EPOS~3 model was used for this purpose in the 1D analysis. The systematic uncertainty associated with the baseline was estimated using an alternative MC model, DPMJET, as well as the two methods based on the use of polynomials described in Ref.~\cite{ALICE:2019kno}. In the 3D analysis, a flat baseline and the DPMJET model were considered as alternative baseline descriptions.

The smearing of the single-particle momenta due to momentum resolution reduces the height and increases the width of the correlation function. It was shown that this effect changes the extracted parameters by $<1$\% both in the 1D and in the 3D analyses, estimated as described in Ref.~\cite{ALICE:2022mxo} and in Ref.~\cite{ALICE:2017iga}, respectively.

The Barlow criterion~\cite{Barlow:2002yb} was applied to estimate a statistical significance level of the effect produced by every systematic contribution $i$ as
\begin{eqnarray}
\mathfrak{B}=\frac{|y_0-y_{\rm var}^i|}{\sqrt{\sigma_0^2+{\sigma_{\rm var}^i}^2-2\rho\sigma_0\sigma_{\rm var}^i}}, \label{eq:Barlow}
\end{eqnarray}
where $\sigma_0$ is the statistical uncertainty of the central value $y_0$ extracted from the analysis with the default selection criteria and fit parameters, $\sigma_{\rm var}^i$ is the statistical uncertainty of the value $y_{\rm var}^i$ obtained with a given variation $i$ of particle selections and parameters of the fit, and $\rho$ characterizes the correlation between $y_0$ and $y_{\rm var}^i$.
The variation $y_{\rm var}^i$ was included in the total systematic uncertainty
\begin{eqnarray}
\Delta_{\rm sys}=\sqrt{\sum_i(\Delta_{\rm sys}^i)^2}, \label{eq:sys}
\end{eqnarray}
where $\Delta_{\rm sys}^i=|y_0-y_{\rm var}^i|$, if $\mathfrak{B}$ in Eq.~(\ref{eq:Barlow}) was larger than unity.

The contributions to the systematic uncertainty of the radii and correlation strengths are given in Table~\ref{tab:systerrKch_1D_sm} for the 1D analysis in six multiplicity and two $k_{\rm T}$ ranges, in Table~\ref{tab:systerrKch_1D_skt} for the analysis in three multiplicity and eight $k_{\rm T}$ ranges, and in Table~\ref{tab:systerrKch_3D} for the 3D analysis.
\begin{table}[ht!]
\centering
\caption{List of contributions to the systematic uncertainty of the 1D femtoscopic radii $R_{\rm inv}$ and correlation strength $\lambda$ in six multiplicity classes: 0--5\%, 5--10\%, 10--20\%, 20--40\%, 40--60\%, and 60--80\% and two pair transverse momentum $k_{\rm T}$ ranges: (0.2--0.5) and (0.5--1.0)~GeV$/c$. The minimum--maximum values in all multiplicity classes and $k_{\rm T}$ ranges are shown.}
 \begin{tabular}{lcc}
 \hline\hline
 & $R_{\rm inv}$ (\%) & $\lambda$ (\%) \\ \hline
 {DCA} & 0.8--2.2 & 1.2--11 \\
 {Number of points in TPC} & 0.5--10 & 1--11 \\
 {$\eta$ range} & 0.5--10 &  3--21 \\
 {$z$ vertex position} & 0.2--10 &  0.3--12 \\
 {Number of mixed events} & 0.4--3 &  1--11 \\
 {Two-track selection} & {1.4--4} & {1.8--12} \\
 {Baseline} & {2.5--6} & 0.4--4 \\
 {Fit range} & {1--7} & 0.5--4 \\
 {Coulomb radius} & {0.8--1.9} & 0--5 \\ \hline\hline
 \end{tabular}
\label{tab:systerrKch_1D_sm}
\end{table}
%
\begin{table}[ht!]
\centering
\caption{List of contributions to the systematic uncertainty of the 1D femtoscopic radii $R_{\rm inv}$ and correlation strength $\lambda$ in three multiplicity classes: 0--20\%, 20--40\%, and 40--90\% and in eight $k_{\rm T}$ ranges: (0.2--0.3), (0.3--0.4), (0.4--0.5), (0.5--0.6), (0.6--0.7), (0.7--0.8), (0.8--1.0), and (1.0--1.3)~GeV$/c$. The minimum--maximum values in all multiplicity classes and $k_{\rm T}$ ranges are shown.
}
\begin{tabular}{lcc}
\hline\hline
& $R_{\rm inv}$ (\%) & $\lambda$ (\%) \\ \hline
{DCA} & 0.5--13 & 1--19 \\
{Number of points in TPC} & 1--14 & 0--18 \\
{$\eta$ range} & 1--13.5 &  2.5--13 \\
{$z$ vertex position} & 0.2--14 &  1--20 \\
{Number of mixed events} & 1.4--14 &  2--13.5 \\
{Two-track selection} & {1.5--15} & {2--27} \\
{Baseline} & {2--10.5} & 0.5--7 \\
{Fit range} & {0.2--16} & 0.1--11 \\
{Coulomb radius} & {0.5--2} & 4.5--5.5 \\ \hline\hline
\end{tabular}
\label{tab:systerrKch_1D_skt}
\end{table}
%
\begin{table}[ht!]
\centering
\caption{List of contributions to the systematic uncertainty of the 3D femtoscopic radii $R_{\rm out}$, $R_{\rm side}$, $R_{\rm long}$ and correlation strength $\lambda$ in three multiplicity classes: 0--20\%, 20--40\%, and 40--90\% and two pair transverse momentum $k_{\rm T}$ ranges: (0.2--0.5) and (0.5--1.0)~GeV$/c$. The minimum--maximum values in all multiplicity classes and $k_{\rm T}$ ranges are shown.
}
\begin{tabular}{lcccc}
\hline\hline
& $R_{\rm out}$ (\%) & $R_{\rm side}$ (\%) & $R_{\rm long}$ (\%) & $\lambda$ (\%) \\ \hline
{DCA} & 0.2--2.7 & 1.3--9 & 1--4 & 1--2.5 \\
{Number of points in TPC} & 2--9 & 2--5 & 1--5 & 1.5--10 \\
{$\eta$ range} & 2--7.5 &  2--6 & 1--10 &  1--11 \\
{$z$ vertex position} & 0.1--6.5 &  0.2--3.2 & 0.3--5.5 &  0.3--8.5 \\
{Number of mixed events} & 0--6.5 &  0--6 & 1--6 &  0.2--8 \\
{Two-track selection} & {0.2--9} & {0.2--3.5} & {0--6} & {0--10} \\
{Baseline} & {3--16} & 1.5--25 & {1--24} & 6--22 \\
{Fit range} & {1.5--9} & 2--5 & {1.5--7} & 1.3--4 \\
{Coulomb radius} & {1.2--2.1} & 0.8--1.9 & {0.8--1.8} & 4.8--5.2 \\ \hline\hline
\end{tabular}
\label{tab:systerrKch_3D}
\end{table}
%
Selected according to the Barlow criterium, the main sources of systematic uncertainty on the extracted parameters are the baseline description, the single particle and pair selection criteria, and the number of events used to construct the reference distribution $B({\bf q})$ in Eq.~(\ref{eq:CFdef}).

\section{Results and discussion}
\label{sec:results}
\subsection{1D analysis results}
\begin{figure}[ht!]
\centering
\subfigure{\label{fig:RL_smA}\includegraphics[width=0.49\linewidth]{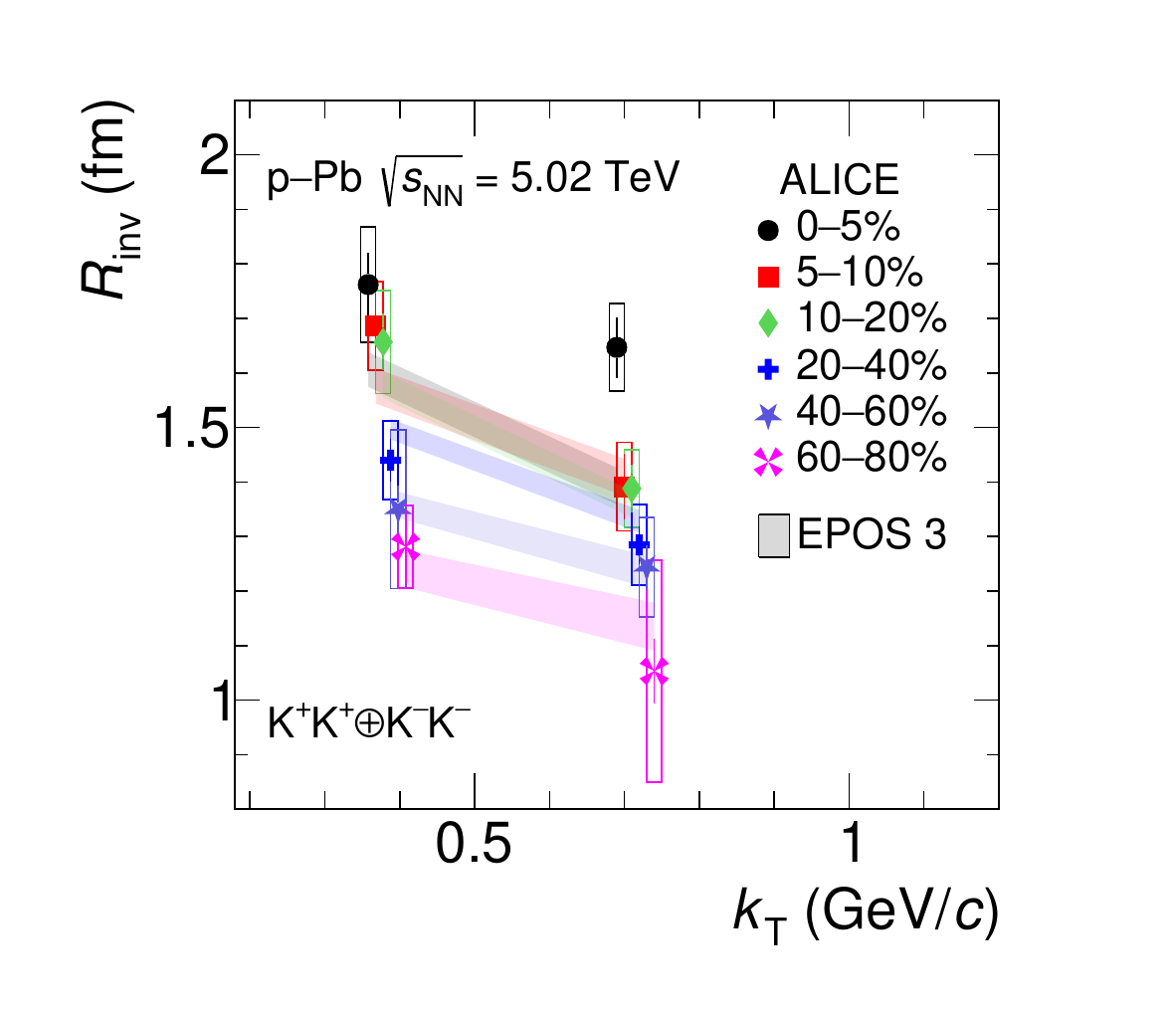}}
\subfigure{\label{fig:RL_smB}\includegraphics[width=0.49\linewidth]{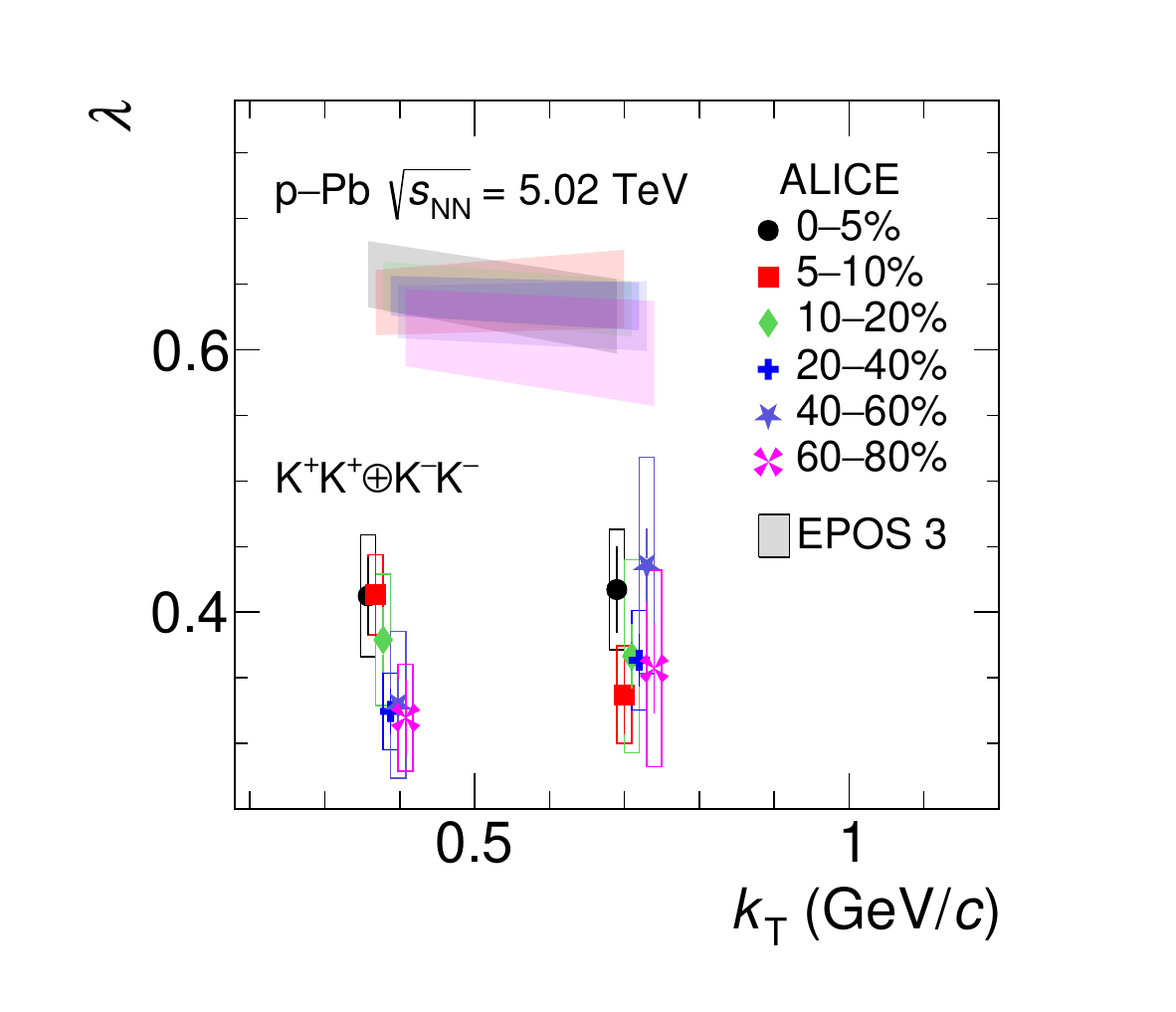}}
\caption{(Color online) Experimental (symbols) K$^\pm$K$^\pm$ invariant radii $R_\mathrm{inv}$ (left) and correlation strengths $\lambda$ (right) shown as a function of the pair transverse momentum $k_{\mathrm T}$ for six multiplicity classes and compared with the EPOS~3 model predictions. The width of the bands represents the statistical uncertainty of the EPOS~3 predictions.
The different colors of the bands correspond to the respective centralities as given by the ALICE data.
Statistical (lines) and systematic uncertainties (boxes) are shown for the data points. The points for lower multiplicity classes are slightly shifted with respect to the 0--5\% multiplicity class (black symbols) in the $x$ direction for clarity.} \label{fig:RL_sm}
\end{figure}
The extracted $R_\mathrm{inv}$ and $\lambda$ parameters in six multiplicity classes and two $k_{\rm T}$ ranges are depicted in Fig.~\ref{fig:RL_sm}. The presented results are also compared with the EPOS~3 model~\cite{Werner:2010aa,Werner:2011fd,Werner:2010ny} with the hadronic afterburner (UrQMD cascade)~\cite{Knospe:2015nva} included, for the same collision system and energy in the same multiplicity and $k_{\rm T}$ ranges as the experimental data. In the previous work on p--Pb collisions~\cite{ALICE:2019kno}, the experimental radii were compared with the EPOS~3 calculations in wider 0--20\%, 20--40\%, and 40--90\% multiplicity classes since the available data at the time did not enable a study in finer multiplicity intervals. The EPOS~3 radii were found to agree with the data in Ref.~\cite{ALICE:2019kno} within uncertainties. The present 1D analysis shows that the radii are well reproduced by the EPOS~3 model calculations for the semicentral and peripheral collisions. At the same time, the model tends to underestimate $R_{\rm inv}$ for the 0--5\% multiplicity interval. The EPOS~3 correlation strength values are about 0.65 and larger than the $\lambda$ values extracted from the experimental data, which lie around 0.35.
The value of the $\lambda$ parameter equals unity in the case of a perfectly Gaussian spherical source with pure primary particles emitted from the source chaotically. In reality, a fraction of kaons could be emitted coherently~\cite{ALICE:2013uhj,Akkelin:2001nd}, and also there are kaons from K$^*$ decays (whose decay width $\Gamma\approx$ is 50~MeV) and from other long-lived resonances~\cite{Wiedemann:1996ig}, whose contribution decreases the $\lambda$ value. A similar difference between the $\lambda$ values from the data and from the EPOS~3 predictions was already observed before in Ref.~\cite{ALICE:2019kno} and was explained by insufficient accounting for all contributions of kaons from various resonance decays in the model.
\begin{figure}[ht!]
\centering
\subfigure{\label{fig:RL_sktA}\includegraphics[width=0.49\linewidth]{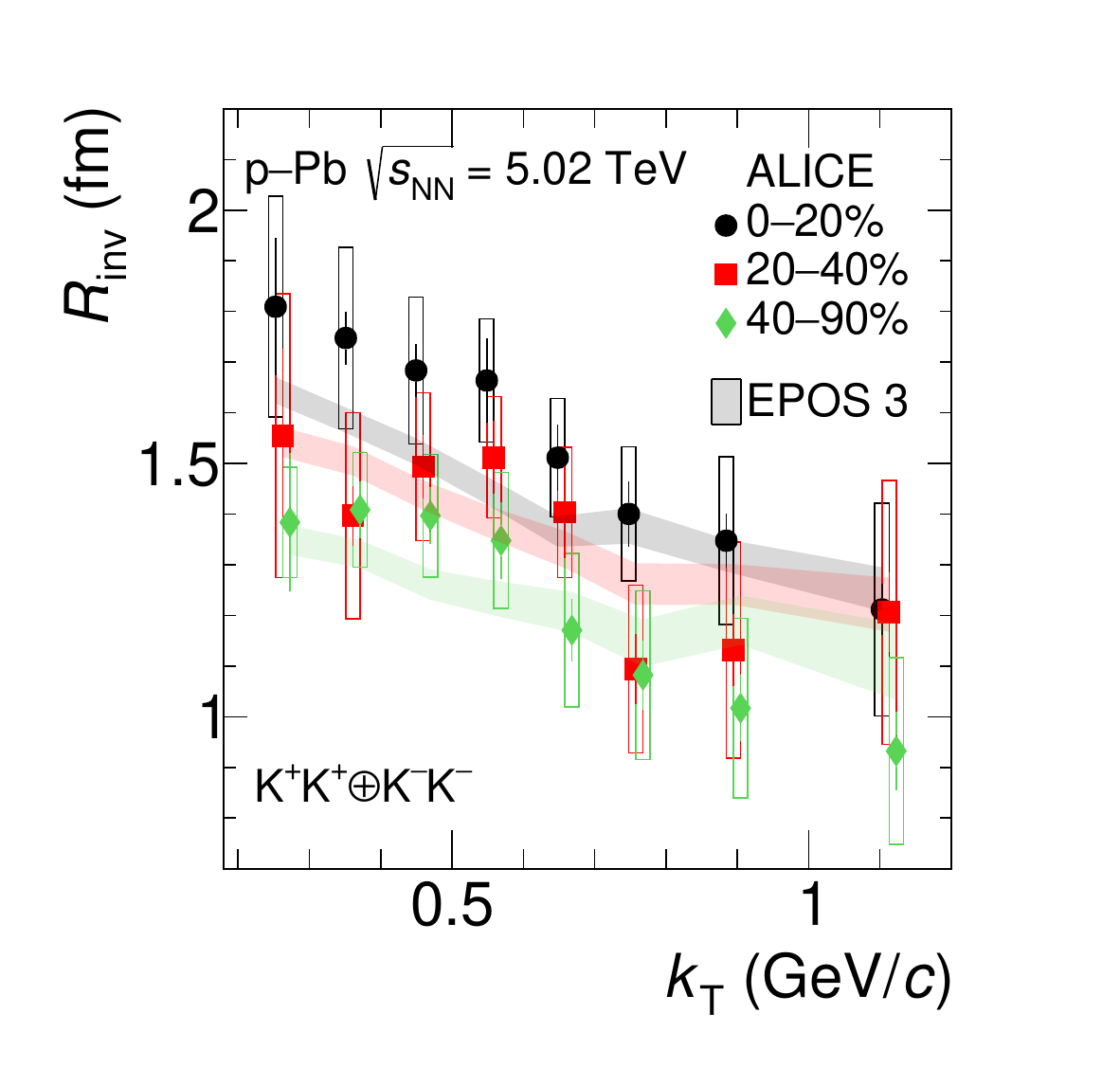}}
\subfigure{\label{fig:RL_sktB}\includegraphics[width=0.49\linewidth]{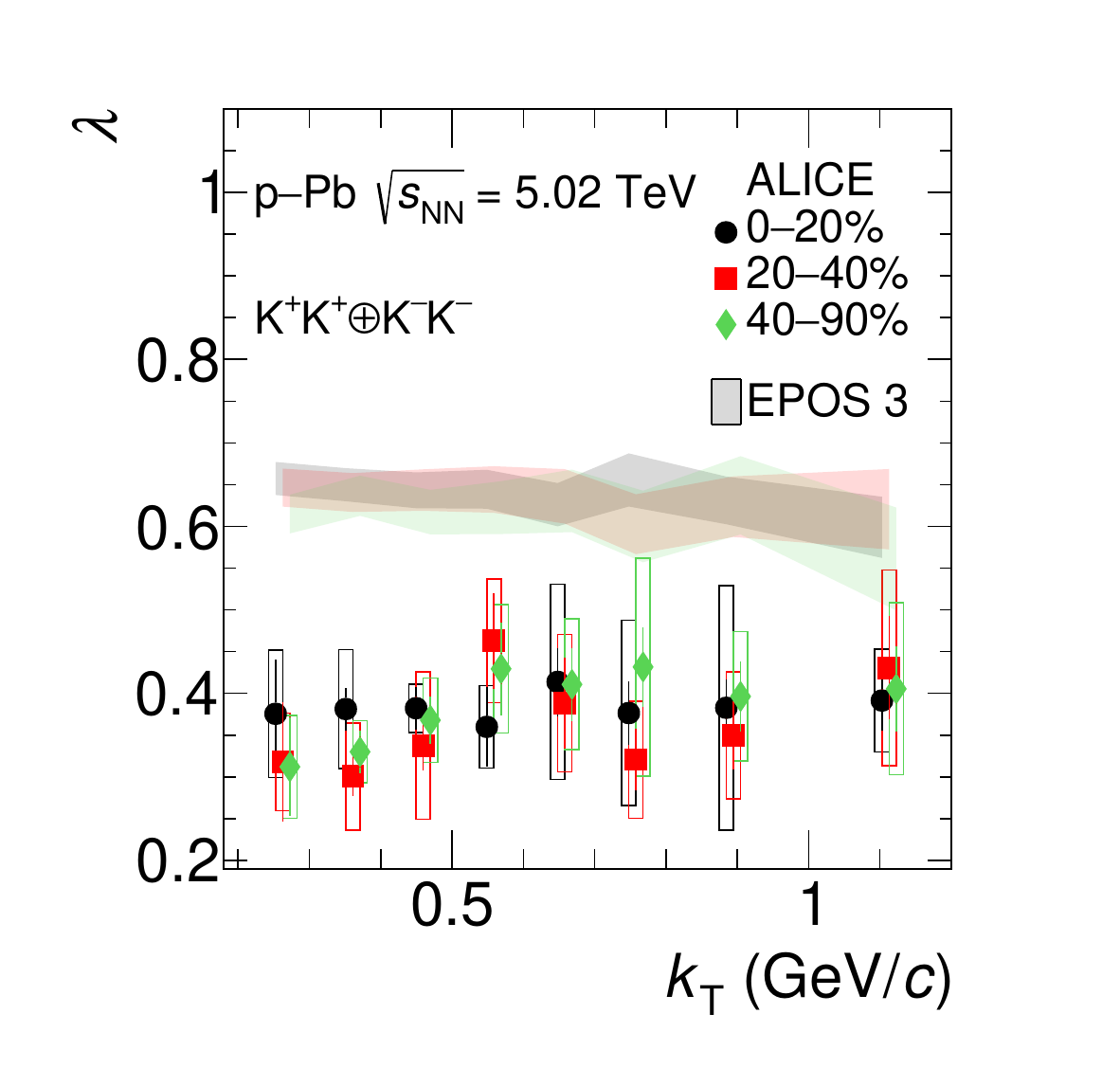}}
\caption{(Color online) Experimental (symbols) K$^\pm$K$^\pm$ invariant radii $R_\mathrm{inv}$ (left) and correlation strengths $\lambda$ (right) shown as a function of the pair transverse momentum $k_{\mathrm T}$ for three multiplicity classes and compared with the EPOS~3 model predictions. The width of the bands represents the statistical uncertainty of the model calculations.
The different colors of the bands correspond to the respective centralities as given by the ALICE data.
Statistical (lines) and systematic uncertainties (boxes) are shown for the data points. The points for lower multiplicity classes are slightly shifted with respect to the 0--20\% multiplicity class (black symbols) in the $x$ direction for clarity.} \label{fig:RL_skt}
\end{figure}

The $R_\mathrm{inv}$ and $\lambda$ parameters in three multiplicity classes and eight $k_{\rm T}$ ranges are shown in Fig.~\ref{fig:RL_skt}, and they provide a more detailed insight into the $k_{\rm T}$ trend. The radii are well reproduced by the EPOS~3 model except for the 0--20\% multiplicity class at low $k_{\rm T}\lesssim0.6$~GeV$/c$ where there is an indication that the model underestimates the radii extracted from the experimental data. This provides a more detailed insight into how the EPOS~3 model describes the data as compared to the previous analysis of p--Pb collisions that was performed in wider $k_{\rm T}$ ranges~\cite{ALICE:2019kno}. For the $\lambda$ parameter, the discrepancy between the EPOS~3 result and the data is similar to that observed in the case of six multiplicity classes and two $k_{\rm T}$ ranges discussed above.

\begin{figure}[ht!]
\centering
\subfigure{\label{fig:Rinv_NchA}\includegraphics[width=0.49\textwidth]{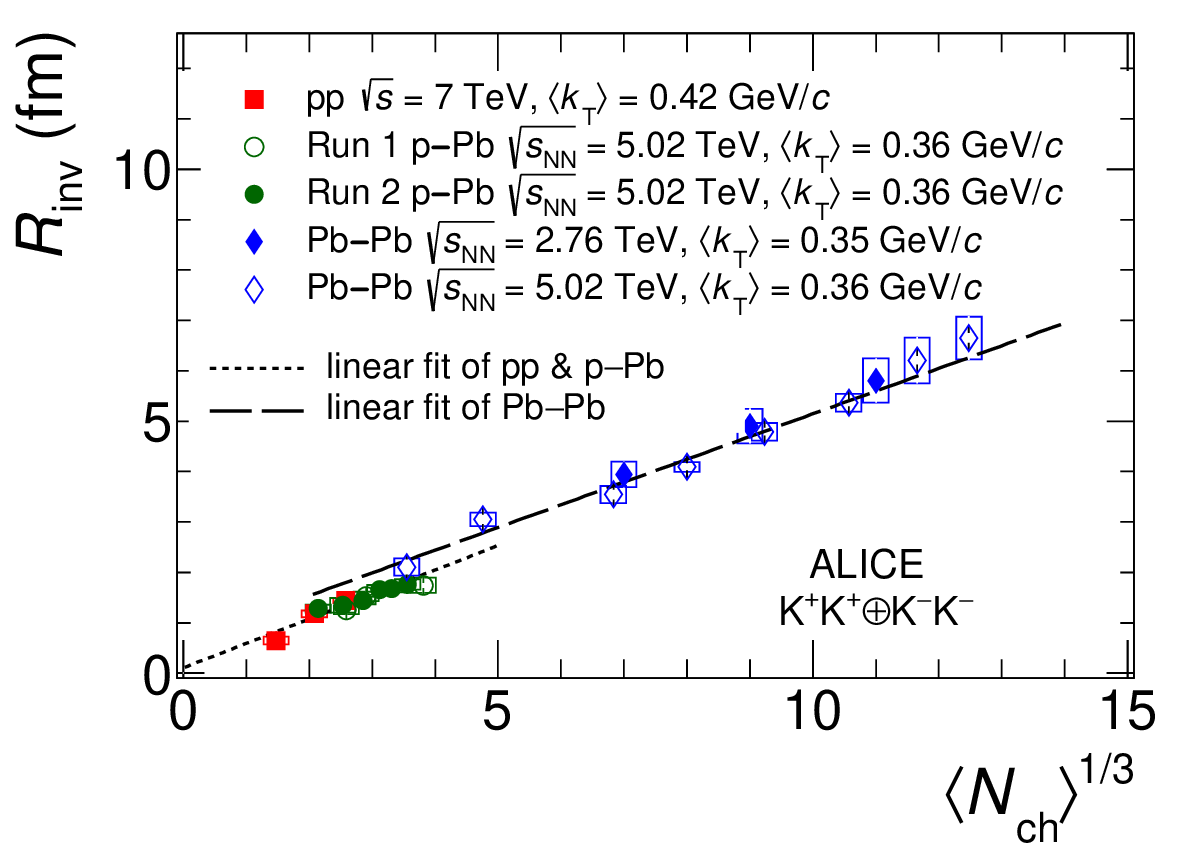}}~~~~~
\subfigure{\label{fig:Rinv_NchB}\includegraphics[width=0.49\textwidth]{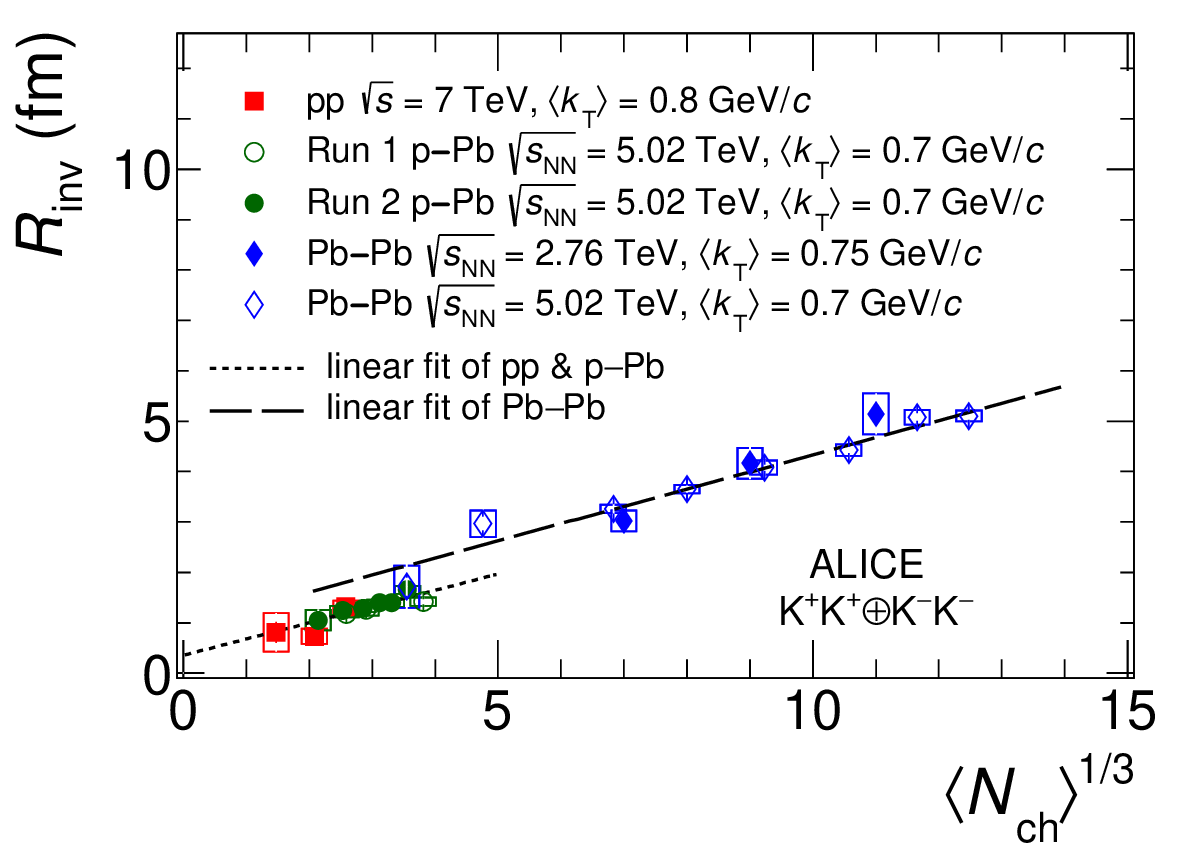}
}
\caption{(Color online) Femtoscopic radii extracted in the present analysis (green solid circles), as a function of the cube root of the measured charged-particle multiplicity $\langle N_{\rm ch}\rangle^{1/3}$, at low (left) and high (right) $k_{\rm T}$ compared with published results from pp~\cite{ALICE:2012aai} (red solid squares), p--Pb~\cite{ALICE:2019kno} (green empty circles), and Pb--Pb~\cite{ALICE:2015hvw,GRomanenko_AN} (blue diamonds) collisions. Statistical (lines) and systematic uncertainties (boxes) are shown. The dotted (dashed) line shows the fit of the pp and p--Pb (Pb--Pb) data with a first-order polynomial.}
\label{fig:Rinv_Nch}
\end{figure}

Figure~\ref{fig:Rinv_Nch} compares the femtoscopic radii extracted in this analysis as a function of the cube root of the measured charged-particle multiplicity $\langle N_{\rm ch}\rangle^{1/3}$, at low (left) and high (right) $k_{\rm T}$ with those obtained in pp~\cite{ALICE:2012aai}, p--Pb~\cite{ALICE:2019kno}, and Pb--Pb~\cite{ALICE:2015hvw,GRomanenko_AN} collisions. The radii increase with $N_{\rm ch}$, and the Run~2 p--Pb results agree within uncertainties with the Run~1 ones~\cite{ALICE:2019kno}. Thanks to the Pb--Pb analysis in the very peripheral centrality ranges~\cite{GRomanenko_AN}, it became possible to compare the radii in p--Pb and Pb--Pb collisions at comparable multiplicities $N_{\rm ch}$. The radii are equal in p--Pb and pp collisions at similar multiplicity within uncertainties. As the figure shows, the trends of the radii as a function of $\langle N_{\rm ch}\rangle^{1/3}$ in pp and p--Pb collisions and separately in Pb--Pb collisions can be well reproduced by a first-order polynomial, supporting the so-called ``scaling hypothesis''~\cite{Lisa:2005dd} that supposes the proportionality of the total multiplicity to the interferometry volume. It is also seen that the line fitting the pp and p--Pb data is shifted vertically with respect to the one fitting the Pb--Pb points, and this shift increases with increasing $k_{\rm T}$ (see Table~\ref{tab:r1D_fit}). The observed different offsets can be considered as an indication of different transverse flow velocities in matter created in pp (p--Pb) and Pb--Pb collisions, originating from the different initial geometric sizes~\cite{Shapoval:2020nec}, or the different source geometry in the final state~\cite{Lisa:2005dd}.
The obtained result reinforces the conclusion made in the earlier p--Pb analysis~\cite{ALICE:2019kno} about the similarity of dynamics of the source in p--Pb and pp collisions at low multiplicities. According to the models in Refs.~\cite{Bozek:2013df,Shapoval:2020nec}, this means that the expansion is not significantly stronger in p--Pb than in pp collisions.
\begin{table}[ht!]
\centering
\caption{Parameters of the $(a+b\langle N_{\rm ch}^{1/3} \rangle)$ fit of the 1D radii in Fig.~\ref{fig:Rinv_Nch}.}
\begin{tabular}{lccc}
\hline\hline
& $a$ & $b$ & $\chi^2$/NDF$^{{\phantom{1}}^{\phantom{1}}}$ \\ \hline
& \multicolumn{3}{c}{low $k_{\rm T}$} \\ \hline
pp\&p--Pb & 0.11$\pm$0.12 & 0.49$\pm$0.04 & 1.42 \\
Pb--Pb & 0.64$\pm$0.23 & 0.45$\pm$0.03 & 0.83 \\ \hline
& \multicolumn{3}{c}{high $k_{\rm T}$} \\ \hline
pp\&p--Pb & 0.36$\pm$0.19 & 0.32$\pm$0.06 & 1.12 \\
Pb--Pb & 0.93$\pm$0.22 & 0.34$\pm$0.02 & 0.92 \\
\hline\hline
\end{tabular}
\label{tab:r1D_fit}
\end{table}

\subsection{3D analysis results}
\subsubsection{Radius and correlation strength}
\begin{figure}[h!]
\centering
{\includegraphics[width=0.65\linewidth]{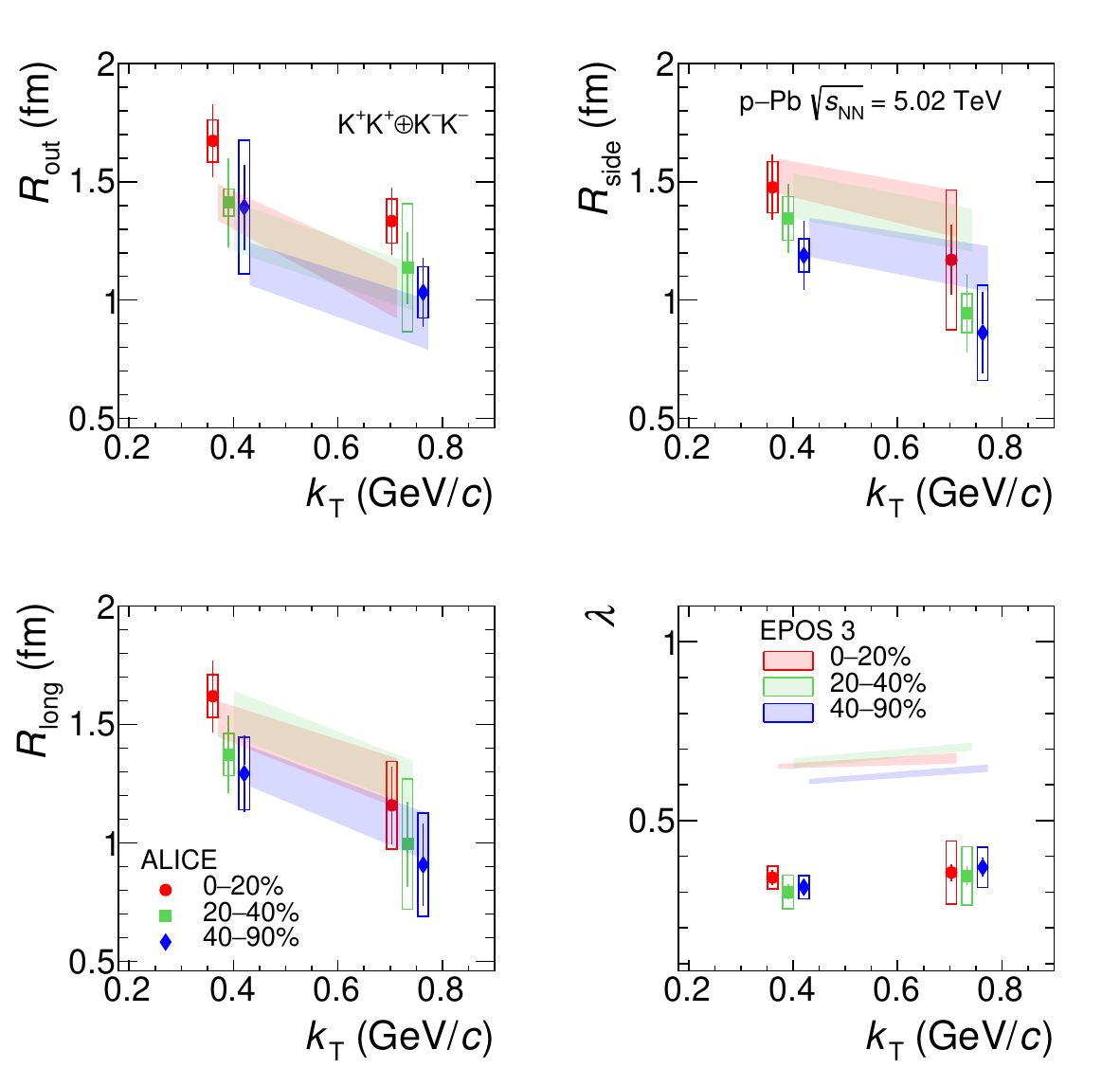}}
\caption{(Color online) 3D $R$ and $\lambda$ parameters as a function of $k_{\rm T}$ in three multiplicity and two $k_{\rm T}$ classes. The experimental data (markers) are compared with the EPOS~3 model predictions shown by bands, the width of which represents the statistical uncertainty on the model calculations. Statistical (bars) and systematic (rectangles) uncertainties are shown. Points are shifted with respect to the 0--20\% multiplicity class along the $x$ axis for clarity.} \label{fig:RoslL_log_sys}
\end{figure}
Figure~\ref{fig:RoslL_log_sys} shows the $k_{\rm T}$ dependence of the extracted 3D kaon radii $R_{\rm out}$, $R_{\rm side}$, $R_{\rm long}$, and correlation strength $\lambda$ in the case of three multiplicity and two $k_{\rm T}$ ranges. As seen from the figure, the extracted experimental radii decrease for more peripheral collisions and with increasing $k_{\rm T}$ within the 1.0 to 1.7~fm interval. The obtained $\lambda$ values do not depend on multiplicity or $k_{\rm T}$, are around $\approx0.3$ and equal within uncertainties to the correlation strength values extracted in the 1D analysis. The EPOS~3 radii agree with the experimental ones within uncertainties, although there is an indication that EPOS~3 underestimates the $R_{\rm out}$ component of the 3D radius for the most central collisions. A similar behavior was observed for K$^\pm$K$^\pm$ correlations in Pb--Pb collisions at $\sqrt{s_{\rm NN}}=5.02$~TeV~\cite{GRomanenko_AN} in comparison to the integrated hydrokinetic model~\cite{Shapoval:2020nec} and for $\pi$K correlations in Pb--Pb collisions at $\sqrt{s_{\rm NN}}=2.76$~TeV~\cite{ALICE:2020mkb} in comparison to the (3+1)D hydrodynamic model~+~THERMINATOR~2~\cite{Kisiel:2018wie}. Therefore, to provide a better description of the femtoscopic data, hydrodynamic models should be improved. The EPOS~3 predictions for the $\lambda$ parameter do not depend on $k_{\rm T}$ and multiplicity and are larger ($\approx0.65$) than the experimental ones similarly to the 1D analysis result, possibly due to the contributions of kaons from long-lived (for example, K$^*$) resonance decays.
\begin{figure}[h!]
\centering
{\includegraphics[width=0.65\linewidth]{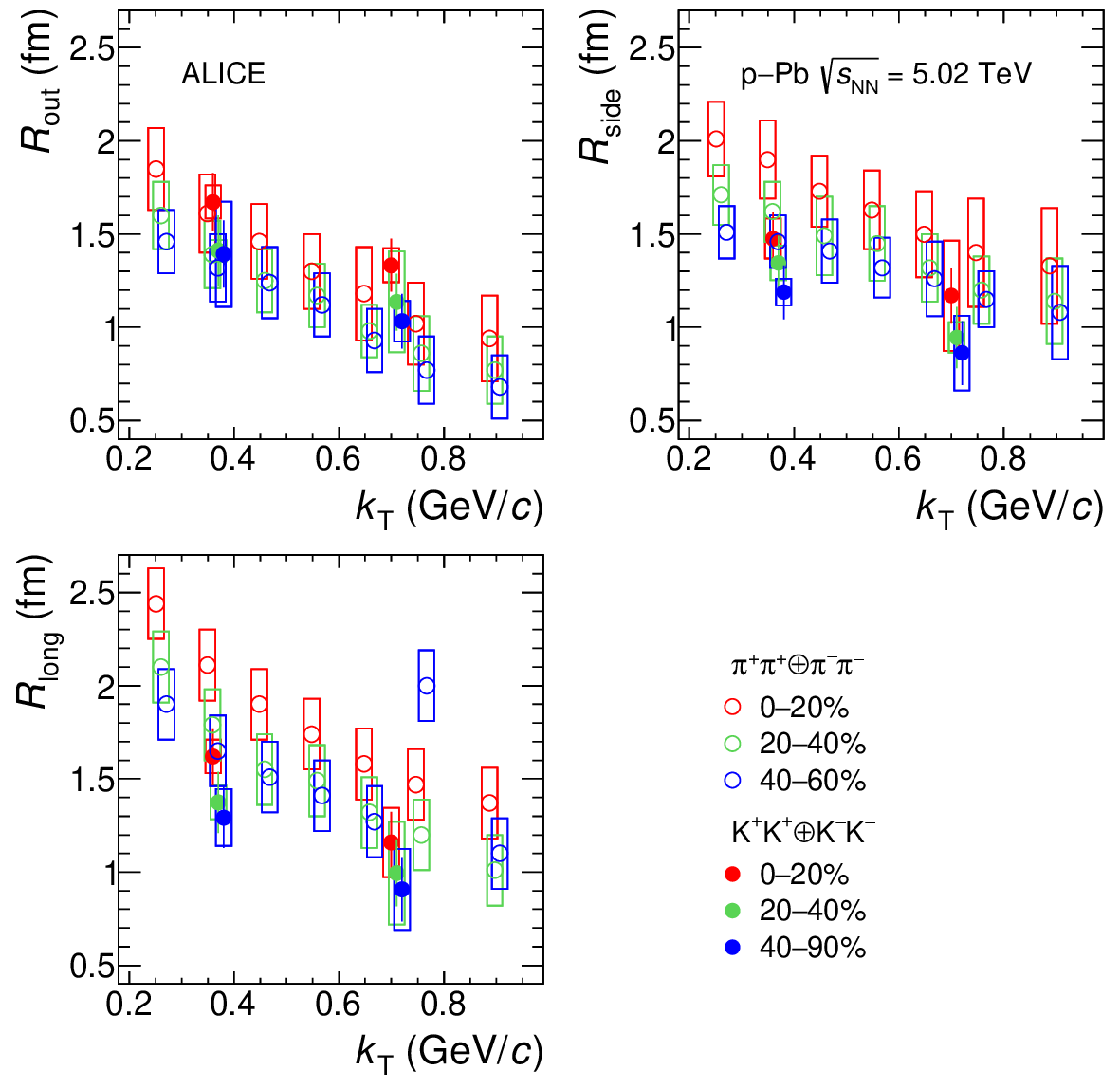}}
\caption{(Color online) 3D $R$ parameters as a function of $k_{\rm T}$ for pion (empty circles) and kaon (full circles) pairs. Statistical (bars) and systematic (rectangles) uncertainties are shown. Points are shifted with respect to the 0--20\% multiplicity class along the $x$ axis for clarity.} \label{fig:RpiK_kt}
\end{figure}

Figure~\ref{fig:RpiK_kt} compares the extracted 3D radii $R_{\rm out}$, $R_{\rm side}$, and $R_{\rm long}$ for pion~\cite{ALICE:2015hav} and kaon pairs as a function of $k_{\rm T}$. The comparison demonstrates that the pion and kaon radii agree within uncertainties at the same multiplicity and $k_{\rm T}$. Unfortunately, the quite large uncertainties of the available experimental data for pions and kaons do not allow for a fine comparison of the radii to see if they scale better with the pair transverse mass $m_{\rm T}$ or the pair transverse momentum $k_{\rm T}$. The $k_{\rm T}$ scaling was observed in the 3D analysis of identical charged kaon correlations in Pb--Pb collisions at $\sqrt{s_{\rm NN}}=2.76$~TeV~\cite{ALICE:2017iga}, although the hydrodynamic scenario of nuclear collisions for the particular case of small transverse flow predicted the $m_{\rm T}$ scaling for the longitudinal radii ($R_{\rm long}$) for pions and kaons~\cite{Makhlin:1987gm}. The obtained $k_{\rm T}$ scaling instead was interpreted as an indication of the importance of the hadronic rescattering phase at LHC energies~\cite{PHENIX:2015jaj,Karpenko:2012yf,Karpenko:2010te}. However, more experimental data are needed to clarify whether $m_{\rm T}$ or $k_{\rm T}$ scaling occurs for pions and kaons in p--Pb collisions.

\subsubsection{Time of maximal emission}
In this work, based on the combined fit of the kaon $R^2_{\rm long}$ dependence on $m_{\rm T}$ and the pion and kaon transverse momentum spectra, the time of maximal emission~\cite{Sinyukov:2015kga,Shapoval:2020nec} for kaons $\tau_{\rm K}$ was estimated as follows.

The first step is to perform a combined fit of pion and kaon $p_{\rm T}$ spectra using
\begin{eqnarray}
 E\frac{{\rm d}^3N}{{\rm d}^3p} \propto \exp\left[-\left(\frac{m_{\rm T}}{T}+\alpha\right)\sqrt{1-\bar{\nu}^2_{\rm T}}\right],\label{eq:spectra_fit}
\end{eqnarray}
where $T$ is the temperature of maximal emission, $\alpha$ is a parameter characterizing the strength of collective flow (infinite value means zero flow, and small value means strong flow), and $\bar{\nu}_{\rm T}=\frac{k_{\rm T}}{m_{\rm T}+\alpha T}$ is the transverse collective velocity. From the fit, the common effective temperature of pions and kaons $T$ and the value of $\alpha$ for each particle species are extracted.

At the second step, the fit of $R^2_{\rm long}$ for kaons is carried out using the formula
\begin{eqnarray}
 &R^2_{\rm long}(m_{\rm T}) = \tau^2 \zeta^2 \left(1+\frac{3}{2}\zeta^2\right),&\label{eq:R_tau}\\
 &\zeta^2 = \frac{T}{m_{\rm T}}\sqrt{1-\bar{\nu}^2_{\rm T}}.&\label{eq:lambda_tau}
\end{eqnarray}
When performing this fit, the temperature $T$ is taken from the fit of the spectra. However, the $\alpha_{\rm K}$ parameter for kaons was
left unconstrained following the approach used for A--A collisions~\cite{Sinyukov:2015kga,Shapoval:2020nec,Shapoval:2021fqg}.

The $\tau_{\rm K}$ parameter was also extracted in the femtoscopic identical charged kaon analyses in Pb--Pb collisions at $\sqrt{s_{\rm NN}}=2.76$~TeV~\cite{ALICE:2017iga} and at $\sqrt{s_{\rm NN}}=5.02$~TeV~\cite{GRomanenko_AN}. Unlike the mentioned Pb--Pb analyses, the available experimental data in p--Pb are limited. Therefore, in this work, a simultaneous fit of pion and kaon $p_{\rm T}$ spectra (Eq.~(\ref{eq:spectra_fit})) and kaon $R_{\rm long}$ (Eq.~(\ref{eq:R_tau})) was performed to extract the time of kaon maximal emission $\tau_{\rm K}$. Such a fitting procedure provided the stability of the extracted $\tau_{\rm K}$ value.
The spectra were taken from Ref.~\cite{ALICE:2013wgn}. In addition, since the three $\tau_{\rm K}$ values extracted in the three corresponding multiplicity classes were found to agree within uncertainties, they were averaged together to obtain a single $\tau_{\rm K}$ value with a smaller statistical uncertainty. The systematic uncertainties of $\tau_{\rm K}$ were determined by propagating the systematic uncertainties of $R_{\rm long}$.

\begin{figure}[h!]
\centering
{\includegraphics[width=0.65\linewidth]{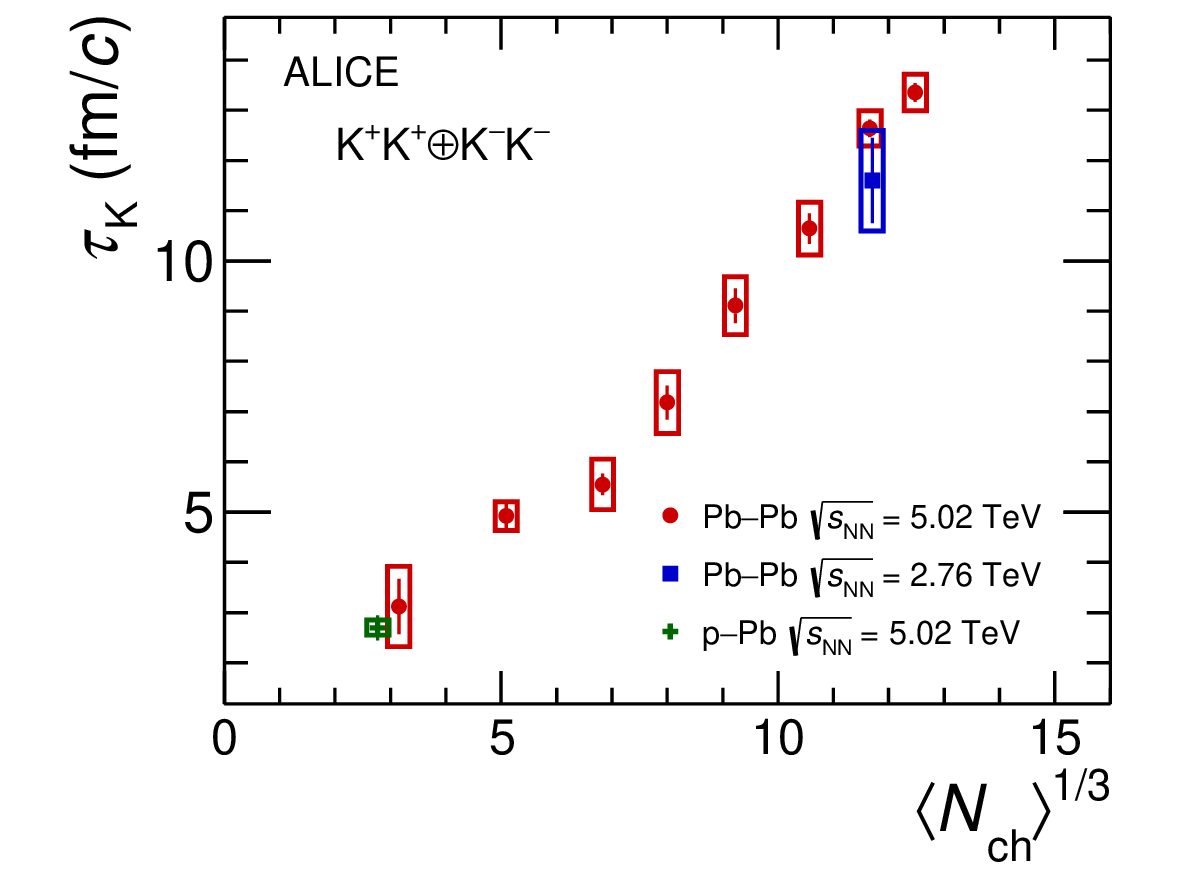}}
\caption{(Color online) Time of maximal emission $\tau_{\rm K}$ as a function of the cube root of the charged particle multiplicity $\langle N_{\rm ch}\rangle^{1/3}$ in p--Pb (green symbol, this work) and Pb--Pb (red~\cite{GRomanenko_AN} and blue~\cite{ALICE:2017iga} symbols) collisions.} \label{fig:tau_nch}
\end{figure}

The resulting $\tau_{\rm K}=2.7\pm0.25{\rm(stat.)}\pm0.15{\rm(syst.)}$ is shown in Fig.~\ref{fig:tau_nch}. It agrees within uncertainties with the value of the time of maximal emission in very peripheral Pb--Pb collisions at the same collision energy. Thus, the kaon emission duration in p--Pb collisions is comparable with that in Pb--Pb collisions at a similar multiplicity. In terms of physical interpretation, it means that the kaon emission process in p--Pb and in very peripheral Pb--Pb collisions occurs similarly over time.
\section{Summary}
\label{sec:summary}
In summary, identical charged kaon pair 1D and 3D correlations were measured in p--Pb collisions at $\sqrt{s_{\rm NN}}=5.02$~TeV. The source radii decrease with increasing pair transverse mass and decreasing event multiplicity both in the 1D and the 3D analyses, similar to the measurements in A--A and high-multiplicity pp collisions in accordance with the hydrodynamic expansion scenario of the medium created in high-energy collisions. The comparison of the obtained results with the EPOS~3 predictions shows that this model does not describe $R_{\rm inv}$ for the most central events, while the description is very good for semicentral and peripheral collisions. The model describes, within uncertainties, the measured 3D radii for all the multiplicity classes and $k_{\rm T}$ ranges, although there is an indication of discrepancy between the experimental and the model $R_{\rm out}$. However, in order to make a statistically significant statement for this discrepancy more experimental data are needed. The comparison of $R_{\rm inv}$ as a function of the multiplicity in pp, p--Pb, and Pb--Pb collisions demonstrates that pp and p--Pb points follow the same linear trend. The linear trend for the $R_{\rm inv}$ values in Pb--Pb collisions has a similar slope but a different offset. The gap between these two fitting lines tends to increase with increasing $k_{\rm T}$. The same observation was done for the pion correlation analysis earlier. This result disfavors models that incorporate substantially stronger collective expansion in p--Pb collisions compared to pp collisions at similar multiplicity.

The correlation strength $\lambda$ is independent of multiplicity or $k_{\rm T}$ within uncertainties. The values of the $\lambda$ parameters calculated with EPOS~3 are larger than those extracted from the experimental data. The observed discrepancy could be due to the influence of long-lived resonances.

The 3D pion and kaon radii agree within uncertainties at the same multiplicity and pair transverse momentum $k_{\rm T}$. The available experimental data do not allow concluding whether $\pi^\pm\pi^\pm$ and K$^\pm$K$^\pm$ radii scale with the transverse mass or the transverse momentum.

Based on the 3D analysis, the time of maximal emission $\tau_{\rm K}$ was extracted for identical charged kaons in p--Pb collisions. The obtained value is equal to the value of $\tau_{\rm K}$ in very peripheral Pb--Pb collisions at the same collision energy indicating the same kaon emission duration in the matter created in p--Pb and Pb--Pb collisions at similar multiplicity.


\newenvironment{acknowledgement}{\relax}{\relax}
\begin{acknowledgement}
\section*{Acknowledgements}

The ALICE Collaboration would like to thank all its engineers and technicians for their invaluable contributions to the construction of the experiment and the CERN accelerator teams for the outstanding performance of the LHC complex.
The ALICE Collaboration gratefully acknowledges the resources and support provided by all Grid centres and the Worldwide LHC Computing Grid (WLCG) collaboration.
The ALICE Collaboration acknowledges the following funding agencies for their support in building and running the ALICE detector:
A. I. Alikhanyan National Science Laboratory (Yerevan Physics Institute) Foundation (ANSL), State Committee of Science and World Federation of Scientists (WFS), Armenia;
Austrian Academy of Sciences, Austrian Science Fund (FWF): [M 2467-N36] and Nationalstiftung f\"{u}r Forschung, Technologie und Entwicklung, Austria;
Ministry of Communications and High Technologies, National Nuclear Research Center, Azerbaijan;
Rede Nacional de Física de Altas Energias (Renafae), Financiadora de Estudos e Projetos (Finep), Funda\c{c}\~{a}o de Amparo \`{a} Pesquisa do Estado de S\~{a}o Paulo (FAPESP) and Universidade Federal do Rio Grande do Sul (UFRGS), Brazil;
Bulgarian Ministry of Education and Science, within the National Roadmap for Research Infrastructures 2020-2027 (object CERN), Bulgaria;
Ministry of Education of China (MOEC) , Ministry of Science \& Technology of China (MSTC) and National Natural Science Foundation of China (NSFC), China;
Ministry of Science and Education and Croatian Science Foundation, Croatia;
Centro de Aplicaciones Tecnol\'{o}gicas y Desarrollo Nuclear (CEADEN), Cubaenerg\'{\i}a, Cuba;
Ministry of Education, Youth and Sports of the Czech Republic, Czech Republic;
The Danish Council for Independent Research | Natural Sciences, the VILLUM FONDEN and Danish National Research Foundation (DNRF), Denmark;
Helsinki Institute of Physics (HIP), Finland;
Commissariat \`{a} l'Energie Atomique (CEA) and Institut National de Physique Nucl\'{e}aire et de Physique des Particules (IN2P3) and Centre National de la Recherche Scientifique (CNRS), France;
Bundesministerium f\"{u}r Forschung, Technologie und Raumfahrt (BMFTR) and GSI Helmholtzzentrum f\"{u}r Schwerionenforschung GmbH, Germany;
General Secretariat for Research and Technology, Ministry of Education, Research and Religions, Greece;
National Research, Development and Innovation Office, Hungary;
Department of Atomic Energy Government of India (DAE), Department of Science and Technology, Government of India (DST), University Grants Commission, Government of India (UGC) and Council of Scientific and Industrial Research (CSIR), India;
National Research and Innovation Agency - BRIN, Indonesia;
Istituto Nazionale di Fisica Nucleare (INFN), Italy;
Japanese Ministry of Education, Culture, Sports, Science and Technology (MEXT) and Japan Society for the Promotion of Science (JSPS) KAKENHI, Japan;
Consejo Nacional de Ciencia (CONACYT) y Tecnolog\'{i}a, through Fondo de Cooperaci\'{o}n Internacional en Ciencia y Tecnolog\'{i}a (FONCICYT) and Direcci\'{o}n General de Asuntos del Personal Academico (DGAPA), Mexico;
Nederlandse Organisatie voor Wetenschappelijk Onderzoek (NWO), Netherlands;
The Research Council of Norway, Norway;
Pontificia Universidad Cat\'{o}lica del Per\'{u}, Peru;
Ministry of Science and Higher Education, National Science Centre and WUT ID-UB, Poland;
Korea Institute of Science and Technology Information and National Research Foundation of Korea (NRF), Republic of Korea;
Ministry of Education and Scientific Research, Institute of Atomic Physics, Ministry of Research and Innovation and Institute of Atomic Physics and Universitatea Nationala de Stiinta si Tehnologie Politehnica Bucuresti, Romania;
Ministerstvo skolstva, vyskumu, vyvoja a mladeze SR, Slovakia;
National Research Foundation of South Africa, South Africa;
Swedish Research Council (VR) and Knut \& Alice Wallenberg Foundation (KAW), Sweden;
European Organization for Nuclear Research, Switzerland;
Suranaree University of Technology (SUT), National Science and Technology Development Agency (NSTDA) and National Science, Research and Innovation Fund (NSRF via PMU-B B05F650021), Thailand;
Turkish Energy, Nuclear and Mineral Research Agency (TENMAK), Turkey;
National Academy of  Sciences of Ukraine, Ukraine;
Science and Technology Facilities Council (STFC), United Kingdom;
National Science Foundation of the United States of America (NSF) and United States Department of Energy, Office of Nuclear Physics (DOE NP), United States of America.
In addition, individual groups or members have received support from:
Czech Science Foundation (grant no. 23-07499S), Czech Republic;
FORTE project, reg.\ no.\ CZ.02.01.01/00/22\_008/0004632, Czech Republic, co-funded by the European Union, Czech Republic;
European Research Council (grant no. 950692), European Union;
ICSC - Centro Nazionale di Ricerca in High Performance Computing, Big Data and Quantum Computing, European Union - NextGenerationEU;
Academy of Finland (Center of Excellence in Quark Matter) (grant nos. 346327, 346328), Finland;
Deutsche Forschungs Gemeinschaft (DFG, German Research Foundation) ``Neutrinos and Dark Matter in Astro- and Particle Physics'' (grant no. SFB 1258), Germany.

\end{acknowledgement}

\bibliographystyle{utphys}   
\bibliography{biblio}

\newpage
\appendix
\section{The ALICE Collaboration}
\label{app:collab}
\begin{flushleft} 
\small

S.~Acharya\,\orcidlink{0000-0002-9213-5329}\,$^{\rm 127}$, 
A.~Agarwal$^{\rm 135}$, 
G.~Aglieri Rinella\,\orcidlink{0000-0002-9611-3696}\,$^{\rm 32}$, 
L.~Aglietta\,\orcidlink{0009-0003-0763-6802}\,$^{\rm 24}$, 
M.~Agnello\,\orcidlink{0000-0002-0760-5075}\,$^{\rm 29}$, 
N.~Agrawal\,\orcidlink{0000-0003-0348-9836}\,$^{\rm 25}$, 
Z.~Ahammed\,\orcidlink{0000-0001-5241-7412}\,$^{\rm 135}$, 
S.~Ahmad\,\orcidlink{0000-0003-0497-5705}\,$^{\rm 15}$, 
S.U.~Ahn\,\orcidlink{0000-0001-8847-489X}\,$^{\rm 71}$, 
I.~Ahuja\,\orcidlink{0000-0002-4417-1392}\,$^{\rm 36}$
% ...

\section*{Affiliation Notes}

$^{\rm I}$ Deceased\\
$^{\rm II}$ Also at: Max-Planck-Institut fur Physik, Munich, Germany\\
$^{\rm III}$ Also at: Czech Technical University in Prague (CZ)\\
$^{\rm IV}$ Also at: Italian National Agency for New Technologies, Energy and Sustainable Economic Development (ENEA), Bologna, Italy\\
$^{\rm V}$ Also at: Dipartimento DET del Politecnico di Torino, Turin, Italy\\
$^{\rm VI}$ Also at: Department of Applied Physics, Aligarh Muslim University, Aligarh, India\\
$^{\rm VII}$ Also at: Institute of Theoretical Physics, University of Wroclaw, Poland\\
$^{\rm VIII}$ Also at: Facultad de Ciencias, Universidad Nacional Aut\'{o}noma de M\'{e}xico, Mexico City, Mexico\\

\section*{Collaboration Institutes}

$^{1}$ A.I. Alikhanyan National Science Laboratory (Yerevan Physics Institute) Foundation, Yerevan, Armenia\\
$^{2}$ AGH University of Krakow, Cracow, Poland\\
$^{3}$ Bogolyubov Institute for Theoretical Physics, National Academy of Sciences of Ukraine, Kyiv, Ukraine\\
$^{4}$ Bose Institute, Department of Physics  and Centre for Astroparticle Physics and Space Science (CAPSS), Kolkata, India\\
$^{5}$ California Polytechnic State University, San Luis Obispo, California, United States\\
$^{6}$ Central China Normal University, Wuhan, China\\
$^{7}$ Centro de Aplicaciones Tecnol\'{o}gicas y Desarrollo Nuclear (CEADEN), Havana, Cuba\\
$^{8}$ Centro de Investigaci\'{o}n y de Estudios Avanzados (CINVESTAV), Mexico City and M\'{e}rida, Mexico\\
$^{9}$ Chicago State University, Chicago, Illinois, United States\\
$^{10}$ China Nuclear Data Center, China Institute of Atomic Energy, Beijing, China\\
$^{11}$ China University of Geosciences, Wuhan, China\\
$^{12}$ Chungbuk National University, Cheongju, Republic of Korea\\
$^{13}$ Comenius University Bratislava, Faculty of Mathematics, Physics and Informatics, Bratislava, Slovak Republic\\
$^{14}$ Creighton University, Omaha, Nebraska, United States\\
$^{15}$ Department of Physics, Aligarh Muslim University, Aligarh, India\\
$^{16}$ Department of Physics, Pusan National University, Pusan, Republic of Korea\\
$^{17}$ Department of Physics, Sejong University, Seoul, Republic of Korea\\
$^{18}$ Department of Physics, University of California, Berkeley, California, United States\\
$^{19}$ Department of Physics, University of Oslo, Oslo, Norway\\
$^{20}$ Department of Physics and Technology, University of Bergen, Bergen, Norway\\
$^{21}$ Dipartimento di Fisica, Universit\`{a} di Pavia, Pavia, Italy\\
$^{22}$ Dipartimento di Fisica dell'Universit\`{a} and Sezione INFN, Cagliari, Italy\\
$^{23}$ Dipartimento di Fisica dell'Universit\`{a} and Sezione INFN, Trieste, Italy\\
$^{24}$ Dipartimento di Fisica dell'Universit\`{a} and Sezione INFN, Turin, Italy\\
$^{25}$ Dipartimento di Fisica e Astronomia dell'Universit\`{a} and Sezione INFN, Bologna, Italy\\
$^{26}$ Dipartimento di Fisica e Astronomia dell'Universit\`{a} and Sezione INFN, Catania, Italy\\
$^{27}$ Dipartimento di Fisica e Astronomia dell'Universit\`{a} and Sezione INFN, Padova, Italy\\
$^{28}$ Dipartimento di Fisica `E.R.~Caianiello' dell'Universit\`{a} and Gruppo Collegato INFN, Salerno, Italy\\
$^{29}$ Dipartimento DISAT del Politecnico and Sezione INFN, Turin, Italy\\
$^{30}$ Dipartimento di Scienze MIFT, Universit\`{a} di Messina, Messina, Italy\\
$^{31}$ Dipartimento Interateneo di Fisica `M.~Merlin' and Sezione INFN, Bari, Italy\\
$^{32}$ European Organization for Nuclear Research (CERN), Geneva, Switzerland\\
$^{33}$ Faculty of Electrical Engineering, Mechanical Engineering and Naval Architecture, University of Split, Split, Croatia\\
$^{34}$ Faculty of Nuclear Sciences and Physical Engineering, Czech Technical University in Prague, Prague, Czech Republic\\
$^{35}$ Faculty of Physics, Sofia University, Sofia, Bulgaria\\
$^{36}$ Faculty of Science, P.J.~\v{S}af\'{a}rik University, Ko\v{s}ice, Slovak Republic\\
$^{37}$ Faculty of Technology, Environmental and Social Sciences, Bergen, Norway\\
$^{38}$ Frankfurt Institute for Advanced Studies, Johann Wolfgang Goethe-Universit\"{a}t Frankfurt, Frankfurt, Germany\\
$^{39}$ Fudan University, Shanghai, China\\
$^{40}$ Gangneung-Wonju National University, Gangneung, Republic of Korea\\
$^{41}$ Gauhati University, Department of Physics, Guwahati, India\\
$^{42}$ Helmholtz-Institut f\"{u}r Strahlen- und Kernphysik, Rheinische Friedrich-Wilhelms-Universit\"{a}t Bonn, Bonn, Germany\\
$^{43}$ Helsinki Institute of Physics (HIP), Helsinki, Finland\\
$^{44}$ High Energy Physics Group,  Universidad Aut\'{o}noma de Puebla, Puebla, Mexico\\
$^{45}$ Horia Hulubei National Institute of Physics and Nuclear Engineering, Bucharest, Romania\\
$^{46}$ HUN-REN Wigner Research Centre for Physics, Budapest, Hungary\\
$^{47}$ Indian Institute of Technology Bombay (IIT), Mumbai, India\\
$^{48}$ Indian Institute of Technology Indore, Indore, India\\
$^{49}$ INFN, Laboratori Nazionali di Frascati, Frascati, Italy\\
$^{50}$ INFN, Sezione di Bari, Bari, Italy\\
$^{51}$ INFN, Sezione di Bologna, Bologna, Italy\\
$^{52}$ INFN, Sezione di Cagliari, Cagliari, Italy\\
$^{53}$ INFN, Sezione di Catania, Catania, Italy\\
$^{54}$ INFN, Sezione di Padova, Padova, Italy\\
$^{55}$ INFN, Sezione di Pavia, Pavia, Italy\\
$^{56}$ INFN, Sezione di Torino, Turin, Italy\\
$^{57}$ INFN, Sezione di Trieste, Trieste, Italy\\
$^{58}$ Inha University, Incheon, Republic of Korea\\
$^{59}$ Institute for Gravitational and Subatomic Physics (GRASP), Utrecht University/Nikhef, Utrecht, Netherlands\\
$^{60}$ Institute of Experimental Physics, Slovak Academy of Sciences, Ko\v{s}ice, Slovak Republic\\
$^{61}$ Institute of Physics, Homi Bhabha National Institute, Bhubaneswar, India\\
$^{62}$ Institute of Physics of the Czech Academy of Sciences, Prague, Czech Republic\\
$^{63}$ Institute of Space Science (ISS), Bucharest, Romania\\
$^{64}$ Institut f\"{u}r Kernphysik, Johann Wolfgang Goethe-Universit\"{a}t Frankfurt, Frankfurt, Germany\\
$^{65}$ Instituto de Ciencias Nucleares, Universidad Nacional Aut\'{o}noma de M\'{e}xico, Mexico City, Mexico\\
$^{66}$ Instituto de F\'{i}sica, Universidade Federal do Rio Grande do Sul (UFRGS), Porto Alegre, Brazil\\
$^{67}$ Instituto de F\'{\i}sica, Universidad Nacional Aut\'{o}noma de M\'{e}xico, Mexico City, Mexico\\
$^{68}$ iThemba LABS, National Research Foundation, Somerset West, South Africa\\
$^{69}$ Jeonbuk National University, Jeonju, Republic of Korea\\
$^{70}$ Johann-Wolfgang-Goethe Universit\"{a}t Frankfurt Institut f\"{u}r Informatik, Fachbereich Informatik und Mathematik, Frankfurt, Germany\\
$^{71}$ Korea Institute of Science and Technology Information, Daejeon, Republic of Korea\\
$^{72}$ KTO Karatay University, Konya, Turkey\\
$^{73}$ Laboratoire de Physique Subatomique et de Cosmologie, Universit\'{e} Grenoble-Alpes, CNRS-IN2P3, Grenoble, France\\
$^{74}$ Lawrence Berkeley National Laboratory, Berkeley, California, United States\\
$^{75}$ Lund University Department of Physics, Division of Particle Physics, Lund, Sweden\\
$^{76}$ Marietta Blau Institute, Vienna, Austria\\
$^{77}$ Nagasaki Institute of Applied Science, Nagasaki, Japan\\
$^{78}$ Nara Women{'}s University (NWU), Nara, Japan\\
$^{79}$ National and Kapodistrian University of Athens, School of Science, Department of Physics , Athens, Greece\\
$^{80}$ National Centre for Nuclear Research, Warsaw, Poland\\
$^{81}$ National Institute of Science Education and Research, Homi Bhabha National Institute, Jatni, India\\
$^{82}$ National Nuclear Research Center, Baku, Azerbaijan\\
$^{83}$ National Research and Innovation Agency - BRIN, Jakarta, Indonesia\\
$^{84}$ Niels Bohr Institute, University of Copenhagen, Copenhagen, Denmark\\
$^{85}$ Nikhef, National institute for subatomic physics, Amsterdam, Netherlands\\
$^{86}$ Nuclear Physics Group, STFC Daresbury Laboratory, Daresbury, United Kingdom\\
$^{87}$ Nuclear Physics Institute of the Czech Academy of Sciences, Husinec-\v{R}e\v{z}, Czech Republic\\
$^{88}$ Oak Ridge National Laboratory, Oak Ridge, Tennessee, United States\\
$^{89}$ Ohio State University, Columbus, Ohio, United States\\
$^{90}$ Physics department, Faculty of science, University of Zagreb, Zagreb, Croatia\\
$^{91}$ Physics Department, Panjab University, Chandigarh, India\\
$^{92}$ Physics Department, University of Jammu, Jammu, India\\
$^{93}$ Physics Program and International Institute for Sustainability with Knotted Chiral Meta Matter (WPI-SKCM$^{2}$), Hiroshima University, Hiroshima, Japan\\
$^{94}$ Physikalisches Institut, Eberhard-Karls-Universit\"{a}t T\"{u}bingen, T\"{u}bingen, Germany\\
$^{95}$ Physikalisches Institut, Ruprecht-Karls-Universit\"{a}t Heidelberg, Heidelberg, Germany\\
$^{96}$ Physik Department, Technische Universit\"{a}t M\"{u}nchen, Munich, Germany\\
$^{97}$ Politecnico di Bari and Sezione INFN, Bari, Italy\\
$^{98}$ Research Division and ExtreMe Matter Institute EMMI, GSI Helmholtzzentrum f\"ur Schwerionenforschung GmbH, Darmstadt, Germany\\
$^{99}$ Saga University, Saga, Japan\\
$^{100}$ Saha Institute of Nuclear Physics, Homi Bhabha National Institute, Kolkata, India\\
$^{101}$ School of Physics and Astronomy, University of Birmingham, Birmingham, United Kingdom\\
$^{102}$ Secci\'{o}n F\'{\i}sica, Departamento de Ciencias, Pontificia Universidad Cat\'{o}lica del Per\'{u}, Lima, Peru\\
$^{103}$ SUBATECH, IMT Atlantique, Nantes Universit\'{e}, CNRS-IN2P3, Nantes, France\\
$^{104}$ Sungkyunkwan University, Suwon City, Republic of Korea\\
$^{105}$ Suranaree University of Technology, Nakhon Ratchasima, Thailand\\
$^{106}$ Technical University of Ko\v{s}ice, Ko\v{s}ice, Slovak Republic\\
$^{107}$ The Henryk Niewodniczanski Institute of Nuclear Physics, Polish Academy of Sciences, Cracow, Poland\\
$^{108}$ The University of Texas at Austin, Austin, Texas, United States\\
$^{109}$ Universidad Aut\'{o}noma de Sinaloa, Culiac\'{a}n, Mexico\\
$^{110}$ Universidade de S\~{a}o Paulo (USP), S\~{a}o Paulo, Brazil\\
$^{111}$ Universidade Estadual de Campinas (UNICAMP), Campinas, Brazil\\
$^{112}$ Universidade Federal do ABC, Santo Andre, Brazil\\
$^{113}$ Universitatea Nationala de Stiinta si Tehnologie Politehnica Bucuresti, Bucharest, Romania\\
$^{114}$ University of Cape Town, Cape Town, South Africa\\
$^{115}$ University of Derby, Derby, United Kingdom\\
$^{116}$ University of Houston, Houston, Texas, United States\\
$^{117}$ University of Jyv\"{a}skyl\"{a}, Jyv\"{a}skyl\"{a}, Finland\\
$^{118}$ University of Kansas, Lawrence, Kansas, United States\\
$^{119}$ University of Liverpool, Liverpool, United Kingdom\\
$^{120}$ University of Science and Technology of China, Hefei, China\\
$^{121}$ University of South-Eastern Norway, Kongsberg, Norway\\
$^{122}$ University of Tennessee, Knoxville, Tennessee, United States\\
$^{123}$ University of the Witwatersrand, Johannesburg, South Africa\\
$^{124}$ University of Tokyo, Tokyo, Japan\\
$^{125}$ University of Tsukuba, Tsukuba, Japan\\
$^{126}$ Universit\"{a}t M\"{u}nster, Institut f\"{u}r Kernphysik, M\"{u}nster, Germany\\
$^{127}$ Universit\'{e} Clermont Auvergne, CNRS/IN2P3, LPC, Clermont-Ferrand, France\\
$^{128}$ Universit\'{e} de Lyon, CNRS/IN2P3, Institut de Physique des 2 Infinis de Lyon, Lyon, France\\
$^{129}$ Universit\'{e} de Strasbourg, CNRS, IPHC UMR 7178, F-67000 Strasbourg, France, Strasbourg, France\\
$^{130}$ Universit\'{e} Paris-Saclay, Centre d'Etudes de Saclay (CEA), IRFU, D\'{e}partment de Physique Nucl\'{e}aire (DPhN), Saclay, France\\
$^{131}$ Universit\'{e}  Paris-Saclay, CNRS/IN2P3, IJCLab, Orsay, France\\
$^{132}$ Universit\`{a} degli Studi di Foggia, Foggia, Italy\\
$^{133}$ Universit\`{a} del Piemonte Orientale, Vercelli, Italy\\
$^{134}$ Universit\`{a} di Brescia, Brescia, Italy\\
$^{135}$ Variable Energy Cyclotron Centre, Homi Bhabha National Institute, Kolkata, India\\
$^{136}$ Warsaw University of Technology, Warsaw, Poland\\
$^{137}$ Wayne State University, Detroit, Michigan, United States\\
$^{138}$ Yale University, New Haven, Connecticut, United States\\
$^{139}$ Yonsei University, Seoul, Republic of Korea\\
$^{140}$ Affiliated with an institute formerly covered by a cooperation agreement with CERN\\
$^{141}$ Affiliated with an international laboratory covered by a cooperation agreement with CERN.\\

\end{flushleft} 
\end{document}